\pdfoutput=1
\documentclass[a4paper,11pt]{article}
\usepackage{jheppub}
\usepackage{graphicx}
\usepackage{amsmath}
\usepackage{amsfonts}
\usepackage{xspace}
\usepackage{hyperref}
\usepackage{placeins}
\usepackage{tikz}
\usepackage{tabularx}
\usepackage{multirow}
\usepackage{defs}
\usepackage{subcaption}
\usetikzlibrary{arrows.meta,positioning}
\usepackage{eso-pic}

\preprint{EPA-26-001}
\makeatletter\renewcommand\afterProceedingsSpace{\kern-17pt}\makeatother 

\title{Agentic AI – Physicist Collaboration in Experimental Particle Physics: A Proof-of-Concept Measurement with LEP Open Data}

\author[a]{Anthony Badea,}
\author[c]{Marcello Maggi,}
\author[b]{Christopher McGinn,}
\author[d]{Austin Baty,}
\author[e]{Yi Chen,}
\author[b]{Hannah Bossi,}
\author[b]{Gian Michele Innocenti,}
\author[e]{Jingyu Zhang,}
\author[b]{Yu-Chen Chen,}
\author[b]{Tzu-An Sheng,}
\author[b]{Yen-Jie Lee,}
\author[]{The Electron-Positron Alliance}

\affiliation[a]{Enrico Fermi Institute, University of Chicago, Chicago, IL 60637, USA}
\emailAdd{badea@uchicago.edu}
\affiliation[b]{Laboratory for Nuclear Science, Massachusetts Institute of Technology, Cambridge, MA 02139, USA}
\affiliation[c]{Istituto Nazionale di Fisica Nucleare, Bari Division, 70126 BA, Italy}
\affiliation[d]{Department of Physics, University of Illinois Chicago, Chicago, IL 60607, USA}
\affiliation[e]{Department of Physics, Vanderbilt University, Nashville, TN 37235, USA}%

\abstract{
We present an agentic measurement of the thrust distribution in $e^{+}e^{-}$ collisions at $\sqrt{s}=91.2$~GeV using ALEPH archived data.
This work was carried out through an AI agent–physicist collaboration, using interactive natural-language prompting of Anthropic Claude and OpenAI Codex. Under physicist guidance, the agents wrote and executed all analysis code and drafted the majority of the text included in this paper.
A fully-corrected thrust spectrum is obtained via Iterative Bayesian Unfolding and Monte Carlo based corrections. 
The principal contribution of this work is the demonstration of the agentic analysis, with thrust serving as a benchmark observable for the method. 
Dedicated precision measurements of thrust are reported in the community elsewhere. 
This work represents a step toward a theory--experiment loop in which AI agents assist with experimental measurements and theoretical calculations, and synthesize insights by comparing the results, thereby accelerating the cycle that drives discovery in fundamental physics.
Our work suggests that precision physics, leveraging the open LEP data and advanced theoretical landscape, provides an ideal testing ground for developing AI agent systems for scientific applications.
}

\keywords{quantum chromodynamics, AI agents, electron-positron collisions, event shapes}

\renewcommand\beforetochook{%
  \thispagestyle{empty}%
  \vspace*{\fill}%
  \begin{center}
  \begin{minipage}{0.75\textwidth}
  \centering\itshape
  The authors dedicate this work to the memory of Prof.\ Mauro De Palma (University of Bari, Italy), who passed away recently. He coordinated the Bari group's activities in the ALEPH experiment and championed the use of ALEPH data to explore new research frontiers. This work continues the scientific vision he so enthusiastically supported.
  \end{minipage}
  \end{center}
  \vspace*{\fill}%
  \newpage
  \pagestyle{myplain}\pagenumbering{roman}}

\begin{document}

\eelogo 
\maketitle

\section{Context of Electron-Positron Alliance}
\label{sec:context}

Modern collider datasets provide unprecedented statistical power, and experimental uncertainties on many Standard Model measurements are now at the percent level. In this regime, theoretical uncertainties increasingly limit the overall precision, and those associated with Quantum Chromodynamics (QCD) are among the dominant contributions. To achieve the physics goals of future programs, including the High-Luminosity LHC (HL-LHC) and Future Circular Collider (FCC-ee), theoretical predictions must therefore reach a precision comparable to the expected experimental uncertainties, in many cases at the sub-percent level, so that potential new physics anomalies can be interpreted reliably.

Precision QCD studies can be pursued at hadron colliders such as the LHC, but they are most cleanly and systematically tested in lepton collider environments. At present, however, there is no active high-energy lepton collider operating at Large Electron--Positron collider (LEP) energies or above. To address this, the Electron-Positron Alliance (EPA)~\cite{EEAlliance:website} works with LEP collaboration members to curate and reanalyse archived datasets, opening new opportunities enabled by three decades of theoretical advances and modern experimental techniques. The combination of ultra-precise $e^+e^-$ data and state-of-the-art perturbative and non-perturbative theory tools provides a uniquely powerful environment for high-precision studies. The EPA has produced a series of precision measurements using these archived datasets, including two-particle correlations~\cite{Badea:2019vey}, jet energy spectra and substructure~\cite{Chen:2021uws}, long-range near-side correlations at $WW$ threshold~\cite{Chen:2024lrc}, energy-energy correlators~\cite{Bossi:2025eec}, and an unbinned measurement of thrust~\cite{Badea:2025ubt}.

More broadly, this program is a compelling opportunity for AI for Science in a domain where precision is paramount. While AI is increasingly used to assist theoretical calculations and generate predictions, in the physical sciences such predictions ultimately require experimental validation. In high-energy physics, this validation depends on collider datasets that are rarely publicly accessible. The $e^+e^-$ environment, particularly at LEP energies, provides the cleanest setting for precision tests of the Standard Model (SM), with well-defined initial states and calculable dynamics. The successful curation and use of archived open data therefore establishes the conditions for a theory--experiment loop in which AI systems, guided by domain experts, can propose new calculations, observables, or improved models and directly confront them with high-precision data. This creates a concrete framework for developing AI systems that do not merely interpolate existing knowledge, but actively refine theoretical descriptions through direct comparison with experiment.

\section{Introduction}
\label{sec:introduction}

Machine learning has been central to high-energy physics for decades, enabling
reconstruction algorithms, event classification, and detector optimization across
collider experiments. A new frontier has emerged with large language models (LLMs)
developed by major AI laboratories, including
Anthropic~\cite{Anthropic:2025accelerating}, OpenAI~\cite{OpenAI:2025gpt5science},
Google DeepMind~\cite{GoogleDeepMind:2025geminiscience}, and others, that can now
assist directly with scientific reasoning, mathematical proof, and code generation.
A striking asymmetry persists: AI is transforming the theoretical and computational side of science research, assisting derivations of new results in theoretical particle physics~\cite{OpenAI:2025newresult}, designing novel protein structures for biology and health research~\cite{GoogleDeepMind:2025alphaproteo}, and simplifying complex calculations. Yet the experimental side of the AI-assisted research loop remains largely absent, especially in fundamental physics. To realize a future where AI rapidly accelerates the discovery of new physics, AI-assisted scientific workflows must be developed and validated across both theory and experiment.

A logical first step toward closing this gap is an agentic analysis of real
experimental data with expert physicist oversight. This is precisely the approach
pursued here. We communicated via natural language prompts with the Anthropic Claude and OpenAI Codex agents, brainstormed
analysis goals, then iteratively prompted the AI agents to execute the full
analysis workflow. No analysis code was written by hand. By analogy, the physicist's role is
similar to advising a graduate student, providing expert direction, reviewing
results, and iterating. The agent was supplied with context comprising the published ALEPH thrust paper~\cite{Heister:2003aj}, prior ROOT-based ALEPH analysis macros covering event selection and thrust calculation, internal analysis notes documenting the miniDST data format and detector calibration procedures, and the archived ALEPH data~\cite{ALEPH:docs}. This prompting strategy follows from our understanding in the AI literature that few-shot examples and supplied context improve AI-assisted outputs~\cite{Brown:2020fewshot,Wei:2022cot}. A detailed inventory of the inputs and prompting approach is given in Section~\ref{sec:ai_agent}.

This vision parallels the evolution of AI in neighboring scientific domains. In
materials science, chemistry, and biology, AI work has progressed from computational
prediction with post-hoc human validation toward AI-designed experiments executed
by autonomous robotic systems, creating closed feedback loops where AI proposes and orchestrates experiments and learns from them. High-energy physics faces a parallel opportunity, with exciting early progress for theory calculations~\cite{Abolhasani:2023sdl,MacLeod:2020sdl,Chen:2024scienceagentbench,Lu:2024aiscientist,Woodruff:2026practicalsi,Bubeck:2025coscientist,OpenAITheory:2026singleminus,Schwartz:2026cparameterllm}.
While we cannot deploy robots to construct particle experiments (yet), the emergence of open and curated data can serve as a
virtual laboratory for developing and validating AI-assisted scientific workflows
that operate across the precision theory--experiment
interface~\cite{Badea:2019vey,Chen:2021uws}.  This requires infrastructure on both sides: precise theoretical predictions and experimental measurements with rigorous uncertainty quantification, where AI-assisted reasoning is tested and held accountable to measurable constraints~\cite{Wang:2023aai,Messeri:2024aiu,Woodruff:2026practicalsi}.

Precision QCD is a natural domain for this program, because $e^{+}e^{-}$ annihilation offers both high-precision data and theory calculations. This matters because fine details are consequential: percent-level uncertainties, coupled to rigorous covariance estimates and end-to-end reproducibility, leave little room for an error to pass unnoticed. In hadronic or nuclear collision environments, by contrast, larger uncertainties can more easily conceal mistakes made by an AI system.

From both an AI and precision physics perspective, thrust is an ideal first observable for this program.
Thrust probes both perturbative radiation and non-perturbative dynamics in a
single distribution, is computable at high perturbative order, and has well-established
connections to resummed calculations, making comparison to theory controlled and
interpretable~\cite{PhysRevLett.39.1587,Abbate:2010xh,Heister:2003aj}. Thrust, along with other event-shape observables measured at LEP, continues to anchor comparisons between
perturbative calculations, generator modeling, and non-perturbative hadronization
constraints. Recent thrust-based $\alpha_s$ extractions report
$\alpha_s(m_Z)\simeq 0.1136\pm 0.0012$, below the 2023 world-average value
$\alpha_s(m_Z)=0.1180\pm 0.0009$~\cite{Benitez:2024nav,ParticleDataGroup:2022pth}.
This persistent tension, often referred to as the $e^{+}e^{-}$ $\alpha_{s}$ puzzle, is of great interest because $\alpha_{s}$ is the least precisely measured coupling constant of the Standard Model. The associated uncertainties propagate to all measurements involving QCD, and the effective field theory tools used to extract $\alpha_{s}$ are widely employed throughout the field. Put differently, if something is missed in the simplest case of $e^{+}e^{-}$ collisions, it is likely that similar effects will also be missed in precision studies of more complex systems. Therefore, testing the new AI technology on thrust in $e^+e^-$ collisions opens the door to significant progress in precision physics.

The objective of the agent-physicist collaboration in this work was to measure the thrust spectrum by applying appropriate selections, unfolding, corrections, and uncertainty handling. The agent was asked to execute the full analysis loop: data and simulation processing, unfolding orchestration,
systematic propagation, figure production, and note maintenance with traceable links
to code and outputs. We utilize the thrust measurement as a testing ground to understand the capabilities of the agents, establish a software base to conduct agentic analysis of the LEP data, and identify areas for improvement. We note that areas for improvement include both the core capabilities of the AI and the interaction of the physicist with the AI.
We envision using our work to pursue a next stage, where the same framework
will be extended to additional event-shape observables (C-parameter and heavy-jet
mass), and the approach of using AI-assisted scientific workflows for both theoretical
and experimental analysis will be pursued in a loop between them. 

The remainder of the document is structured as follows. Section~\ref{sec:ai_agent} defines the AI-agent operating model used in this work.
Section~\ref{sec:aleph} summarizes the ALEPH detector. Section~\ref{sec:datasets} describes data and simulation samples.
Section~\ref{sec:event_selection} details reconstruction and selections, including
particle-category and event-level validation. Section~\ref{sec:method} presents
thrust construction and the executable unfolding workflow.
Section~\ref{sec:systematics} discusses uncertainty measurements. Section~\ref{sec:results} presents the results of the measurement. Section~\ref{sec:conclusion} reports the conclusions from the study. 

\section{AI Agent Usage in This Analysis}
\label{sec:ai_agent}

This work includes an explicit evaluation of AI-agent support for an experimental QCD analysis workflow. The central question is not only whether an agent can generate text or isolated scripts, but whether it can execute a full
analysis loop under physicist supervision: data and Monte Carlo (MC) processing, unfolding, systematic propagation, figure production, and note maintenance with traceable links to code and outputs.

\subsection{Scope and Operating Model}

The agents used in this analysis are Anthropic Claude Code and OpenAI Codex, depending on the preference of the physicist. The operating model is human-in-the-loop: the physicist formulates goals and provides expert direction through iterative prompts; the agent executes the corresponding implementation steps; the physicist reviews the outputs and provides feedback; the process is then repeated. Physics decisions remain analyst-owned, while the agent provides implementation speed, consistency checks, and documentation rigor. In practical terms, the agent is used to:
\begin{itemize}
\item write, execute, and monitor analysis code,
\item enforce consistent plotting and note formatting conventions for work communication,
\item maintain an up-to-date understanding of the analysis code and configurations,
\item iterate quickly on ideas and other time-consuming tasks.
\end{itemize}

\noindent Our operating model is intentionally conservative to see how the agent performs when used to accelerate the common physics engineering and documentation tasks, rather than the core scientific decision tasks.

\subsection{Inputs and Context Supplied to Agents}
\label{sec:agent_inputs}

The inputs supplied to the agents at the start of each session were the following.
\begin{itemize}
\item Context: the published ALEPH thrust measurement, prior ROOT-based analysis macros for event selection and thrust calculation, and internal analysis notes documenting the open data format, detector calibration, and object definitions~\cite{Heister:2003aj,ALEPH:docs}.
\item Datasets: the ALEPH LEP1 data files for the 1994 running period at $\sqrt{s}=91.2$~GeV, together with the corresponding ALEPH MC samples. Additional modern MC were given for theoretical uncertainty estimation.
\item Physicist-written prompts: prompts specifying the physics objective (e.g.,\ ``propagate the hadronic-event correction as an independent systematic component with per-bin envelope uncertainties''), acceptance criteria (e.g.,\ ``refolded spectra must reproduce detector-level data within statistical precision''), and output-format expectations (figure style, table layout, note section to update).
\end{itemize}

\subsection{Boundaries, Controls, and Reproducibility}

To keep scientific accountability explicit, each high-impact step has a control point where physicist sign-off is required before results are treated as baseline. The analysis stages where control points were used are: event and object selection definition, unfolding regularization choice, systematic-combination policy, final figure and uncertainty-presentation choices, interpretation description of the work. Beyond these sign-off gates, the physicist performed concrete validation checks at each major analysis stage by inspecting the artifacts of the agents' work. These checks are enumerated here.

\begin{itemize}
    \item Event selection: Cut-flow yields were compared to the published ALEPH numbers~\cite{Heister:2003aj} to verify selection efficiency. Distributions of $\cos\theta_{\mathrm{thrust}}$ and $\varphi$ were inspected for detector artifacts; this step uncovered a localized $\varphi$ excess in the data that was traced to a detector region and checked before proceeding.
    \item Unfolding: Refolded spectra were required to reproduce the detector-level data within statistical precision. An iteration scan confirmed that the refolded-to-measured ratio is stable as the regularization parameter is scanned.
    \item Systematic uncertainties: Each variation's impact on the unfolded distribution was inspected for large bin-to-bin fluctuations. The covariance and correlation matrices for individual components as well as the total were computed.
    \item Results: The unfolded thrust distribution was compared to the ALEPH\,2004 published result~\cite{Heister:2003aj} over the 42~reported bins. Per-bin uncertainty breakdowns and correlation matrices are reported.
    \item All figure captions, table entries, and interpretive statements in this work were reviewed by the physicist for physics accuracy. Agent-generated text was treated as a draft requiring human approval before inclusion in the baseline.
\end{itemize}

Reproducibility is ensured through multiple complementary mechanisms. All analysis steps are version-controlled, with code and configuration managed in a git repository so that any result can be traced to a specific commit and set of inputs. Workflow runs produce structured output artifacts, including summary files that capture the configuration used, the software versions, and caches of the exported spectra and covariance files. Physicist oversight at each review stage provides a layer of verification, with each step requiring explicit physicist approval before being promoted to the nominal result and the analysis to continue. These controls ensure that results are reproducible and that the physicist retains accountability for the scientific claims.

\section{ALEPH Detector}
\label{sec:aleph}

ALEPH was a general-purpose LEP detector with near $4\pi$ coverage and high-precision charged-particle tracking in a 1.5~T solenoidal magnetic field~\cite{Decamp:1990jra,ALEPH:1994ayc, Heister:2003aj}. The detector subsystems most relevant for this event-shape analysis are the central tracker, electromagnetic and hadronic calorimeters, and outer muon chambers.

The central tracking system combined a silicon vertex detector (VDET), an inner tracking chamber (ITC), and a large time projection chamber (TPC) with 21 pad rows covering $|\cos\theta|<0.97$. Together these detectors provided precise track parameters and momentum measurements, and they define the object-quality observables used in this analysis (for example $d_0$, $z_0$, and $N_{\mathrm{TPC}}$, defined in Section~\ref{sec:object_selection}). The transverse-momentum resolution of the combined tracking system is
$\delta p_T/p_T = 6\times10^{-4}\,p_T \oplus 0.005$ (GeV/$c$), the second term arising from
multiple scattering and dominating at low $p_T$.

The calorimeter system comprised two main subsystems. The electromagnetic calorimeter (ECAL) reconstructed isolated photons with a relative energy resolution of $0.18/\sqrt{E\,[\mathrm{GeV}]} + 0.009$, providing the primary response to photons and electromagnetic showers and anchoring the neutral-energy scale for event-shape reconstruction. The hadronic calorimeter (HCAL) with resolution $\sigma_E/E \simeq 0.85/\sqrt{E\,[\mathrm{GeV}]}$ is essential for neutral-hadron energy flow and global visible-energy reconstruction. The muon system outside HCAL supports lepton categorization in the archived ALEPH object definitions.

A key feature of the ALEPH reconstruction is the energy-flow algorithm, which combines tracking and calorimetric information to form a set of energy-flow objects that represent both charged and neutral particle candidates in the event. Tracking information dominates for charged particles, where the momentum resolution is superior to the calorimeter energy resolution. Calorimetric information provides the primary input for neutral particles (photons and neutral hadrons), which leave no track in the TPC. The combined energy-flow reconstruction achieves an energy resolution of $\sigma(E) = (0.59\pm0.03)\sqrt{E/\text{GeV}} + (0.6\pm0.3)$~GeV for hadronic events. This energy-flow approach, used in the ALEPH 2004 published event-shape analysis~\cite{Heister:2003aj}, is the particle-level input adopted for the present analysis. The 1998 ALEPH event-shape analysis~\cite{Barate:1996fi} unfolded to the charged particle only thrust spectrum to minimize model dependence on neutral-energy simulation; our nominal analysis follows the 2004 approach using both charged and neutral energy-flow objects, with a charged-only variation retained as a cross-check.
\section{Data and Simulated Event Samples}
\label{sec:datasets}

The measurement uses archived ALEPH LEP1 data from 1994 at $\sqrt{s}\simeq 91.2$~GeV. This period is used because matched detector-level and generator-level archived MC inputs required for unfolding are available in this configuration. The 1994 sample corresponds to approximately 40~pb$^{-1}$ after quality requirements and provides the largest fully synchronized archived data/MC configuration for this workflow. Table~\ref{tab:sample_bookkeeping} summarizes the event counts in the nominal thrust workflow. The archived MC reco sample contains approximately 752k selected events compared to 1.33M in data. As a result of having fewer simulated events, MC statistical fluctuations are larger than those in data.

\begin{table}[ht!]
\centering
\begin{tabular}{l c c}
\hline
Sample & Total Events & Selected Events \\
\hline
1994 data reco & 1,365,440 & 1,326,376 \\
1994 MC reco & 771,597 & 751,878 \\
1994 MC gen (\texttt{tgen}) & 771,597 & 771,597 \\
1994 MC gen-before (\texttt{tgenBefore}) & 973,769 & 973,769 \\
\hline
\end{tabular}
\caption{Number of total and selected events in the archived data samples.}
\label{tab:sample_bookkeeping}
\end{table}

Detector corrections are derived from archived ALEPH MC processed through the same analysis chain. The archived \textsc{Pythia}~6~\cite{Sjostrand:2000wi} MC was produced with the standard ALEPH detector simulation software chain (\textsc{JETSET}~7.4 parton shower and hadronization with the ALEPH tune, \textsc{GEANT}~3 detector simulation), consistent with the simulation used in the published ALEPH event-shape analyses of the same dataset~\cite{Heister:2003aj}. In the archived sample, the detector-level and generator-level event records are connected through a common event identifier, enabling event-matched response construction and post-unfolding hadronic-selection corrections. 

The unfolding step consumes four synchronized inputs, each a ROOT \texttt{TTree} in the
archived sample: measured data tree (\texttt{t}), detector-level MC tree (\texttt{t}), generator-level MC tree used for response matching (\texttt{tgen}), generator-level MC tree before hadronic event selection (\texttt{tgenBefore}). The ratio \texttt{tgenBefore/tgen} is used to derive a hadronic event-selection correction applied after unfolding. Event matching between reco and gen levels is performed with the unique event identifier. To populate the response matrix for unfolding, generator entries are matched to the reconstructed selected-event identifier set, consistent with the matched-response construction described in Section~\ref{sec:method}.

In addition to the archived baseline MC, alternative modern generator predictions from \textsc{Pythia}~8.317, \textsc{Herwig}~7.1.5, and \textsc{Sherpa}~2.2.6 are used in this work through event-level reweighting variations to estimate theory model dependence. These variations are propagated as dedicated systematic components in the final uncertainty model while preserving the same detector-response baseline.

\section{Event Reconstruction and Selection}
\label{sec:event_selection}

\subsection{Object Definitions}
\label{sec:particle_categories}

The archived reconstruction builds energy-flow candidates from combined
tracking and calorimeter information. Charged particle candidates rely on high-quality tracking
measurements, while neutral particle candidates utilize the calorimeter. The objects are categorized with an integer label \texttt{pwflag} as summarized in Table~\ref{tab:pwflag_definition}.

\begin{table}[ht!]
\centering
\begin{tabular}{c l}
\hline
\texttt{pwflag} & Category \\
\hline
0 & charged hadrons \\
1 & electrons (tracker + ECAL category) \\
2 & muons (tracker + muon category) \\
3 & $V^{0}$ candidates \\
4 & photons \\
5 & neutral hadrons \\
\hline
\end{tabular}
\caption{Object labels used in archived ALEPH events.}
\label{tab:pwflag_definition}
\end{table}

\noindent $V^0$ candidates (\texttt{pwflag}=3) include $K_S$ and $\Lambda$ from displaced-vertex reconstruction. These objects enter the thrust sum with their measured momentum vector rather than individual decay products, which is consistent with treating them as single hadronic objects for event-shape purposes. Their contribution to the thrust numerator and denominator is captured by the neutral-response systematic variations.

\subsection{Object and Event Selections}
\label{sec:object_selection}

Table~\ref{tab:particle_selection_summary} lists the object-level selection. These criteria define which reconstructed objects are propagated into thrust and therefore which detector effects enter the response model. They are imposed on the reconstructed collection only; the generator-level particles of the same events are retained without them. Here $d_0$ and $z_0$ are the transverse and longitudinal impact parameters of the track with respect to the per-event reconstructed primary vertex. Both are taken from the archived reconstruction. $N_{\mathrm{TPC}}$ is the number of TPC coordinates associated with the track.

\begin{table}[ht!]
\centering
\begin{tabular}{@{}p{0.37\textwidth}p{0.57\textwidth}@{}}
\hline
Object Variable & Requirement \\
\hline
Charged category & \texttt{pwflag}\,$\in[0,2]$ \\
Charged acceptance & $|\cos\theta|\le 0.94$ \\
Charged transverse momentum & $p_{\mathrm{T}}\ge 0.2$ GeV \\
Charged impact parameter & $|d_0|\le 2$ cm \\
Charged longitudinal impact & $|z_0|\le 10$ cm \\
Charged hit quality & $N_{\mathrm{TPC}}\ge 4$ \\
Neutral category & \texttt{pwflag}\,$\in[3,5]$ \\
Neutral energy & $E\ge 0.4$ GeV \\
Neutral acceptance & $|\cos\theta|\le 0.98$ \\
Use charged objects & \texttt{keepChargedTracks} = \texttt{True} \\
Use neutral objects & \texttt{keepNeutralTracks} = \texttt{True} \\
\hline
\end{tabular}
\caption{Object selections applied to data and detector level MC.}
\label{tab:particle_selection_summary}
\end{table}

The hadronic event selection is summarized in Table~\ref{tab:event_selection_summary}, following the standard ALEPH selection~\cite{Heister:2003aj, Barate:1996fi}. These selections enforce a phase space with high tracking quality, controlled forward acceptance, and stable visible-energy topology before entering event-shape reconstruction. The residual contamination from processes such as $e^+e^- \rightarrow \tau^+\tau^-$ is expected to be less than 0.26\% for these event selections~\cite{Barate:1996fi}. Approximately 1.36 (1.32) million $e^+e^-$ collisions resulting in the decay of a Z boson to quarks are analyzed (selected). Note that the acceptance requirement based on the event shape variable sphericity introduces a slight sculpting of the thrust distribution. The cut removes approximately 2.3\% of selected events, with the largest effect in the multijet region. In data, the ratio of the thrust distribution with the cut to that without rises monotonically from $\approx 0.85$ at $T \approx 0.6$--$0.75$ to $\approx 0.95$ at $T \approx 0.9$ and $\approx 0.99$ as $T \to 1$.

\begin{table}[ht!]
\centering
\begin{tabular}{@{}p{0.37\textwidth}p{0.57\textwidth}@{}}
\hline
Event Variable & Requirement \\
\hline
Total charged energy & $E_{\mathrm{trk}}^{\mathrm{tot}}\ge 15$ GeV \\
Sphericity-axis acceptance & $|\cos\theta_{\mathrm{Sph}}|\le 0.82$ \\
Charged multiplicity & $N_{\mathrm{trk}}\ge 5$ \\
Total multiplicity & $N_{\mathrm{trk}}+N_{\mathrm{neu}}\ge 13$ \\
\hline
\end{tabular}
\caption{Nominal event selections applied to data and detector level MC to build the nominal thrust input particle set.}
\label{tab:event_selection_summary}
\end{table}


\subsection{Data-MC Kinematic Checks}

\begin{figure}[t!]
\centering
\sfpanel[0.495]{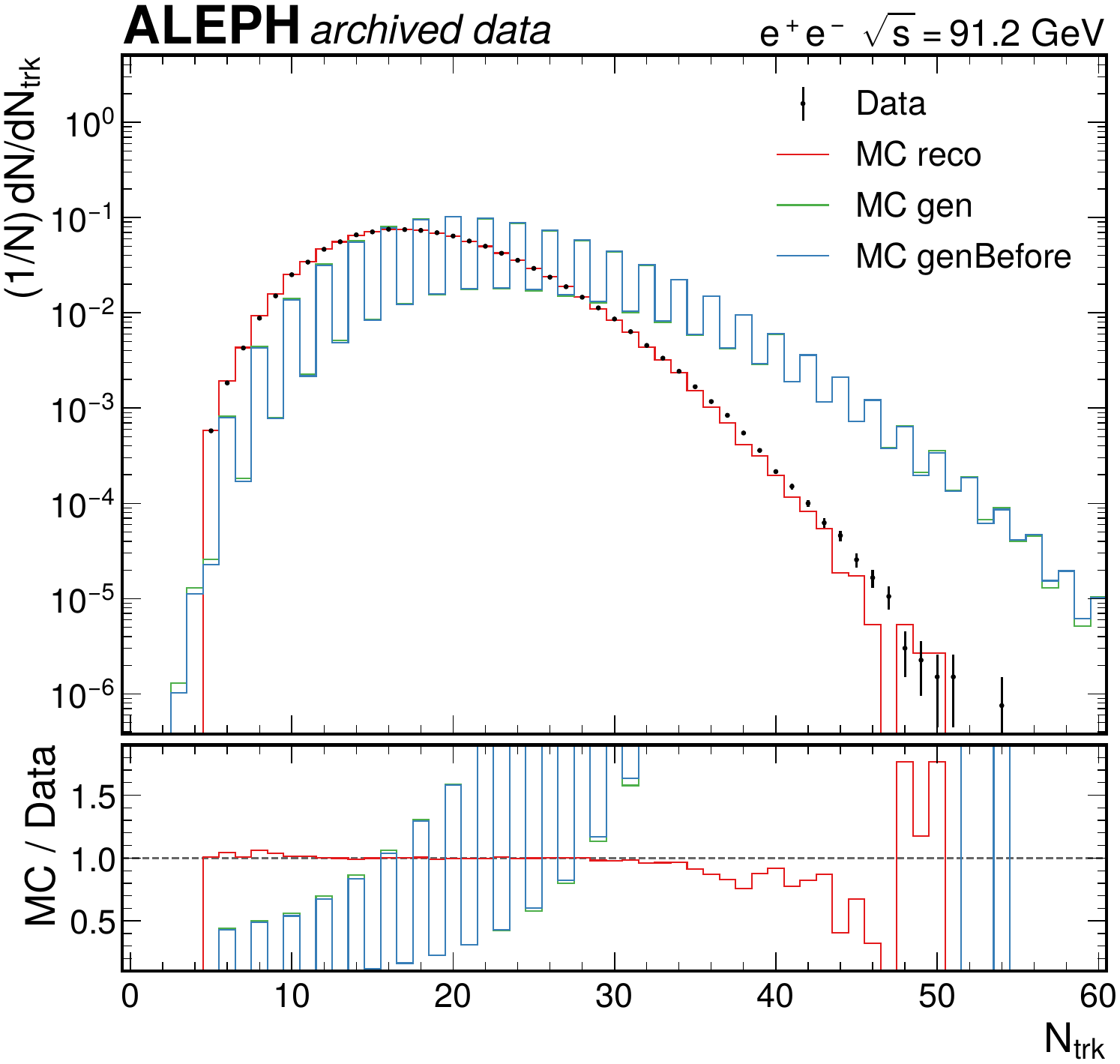}\hfill
\sfpanel[0.495]{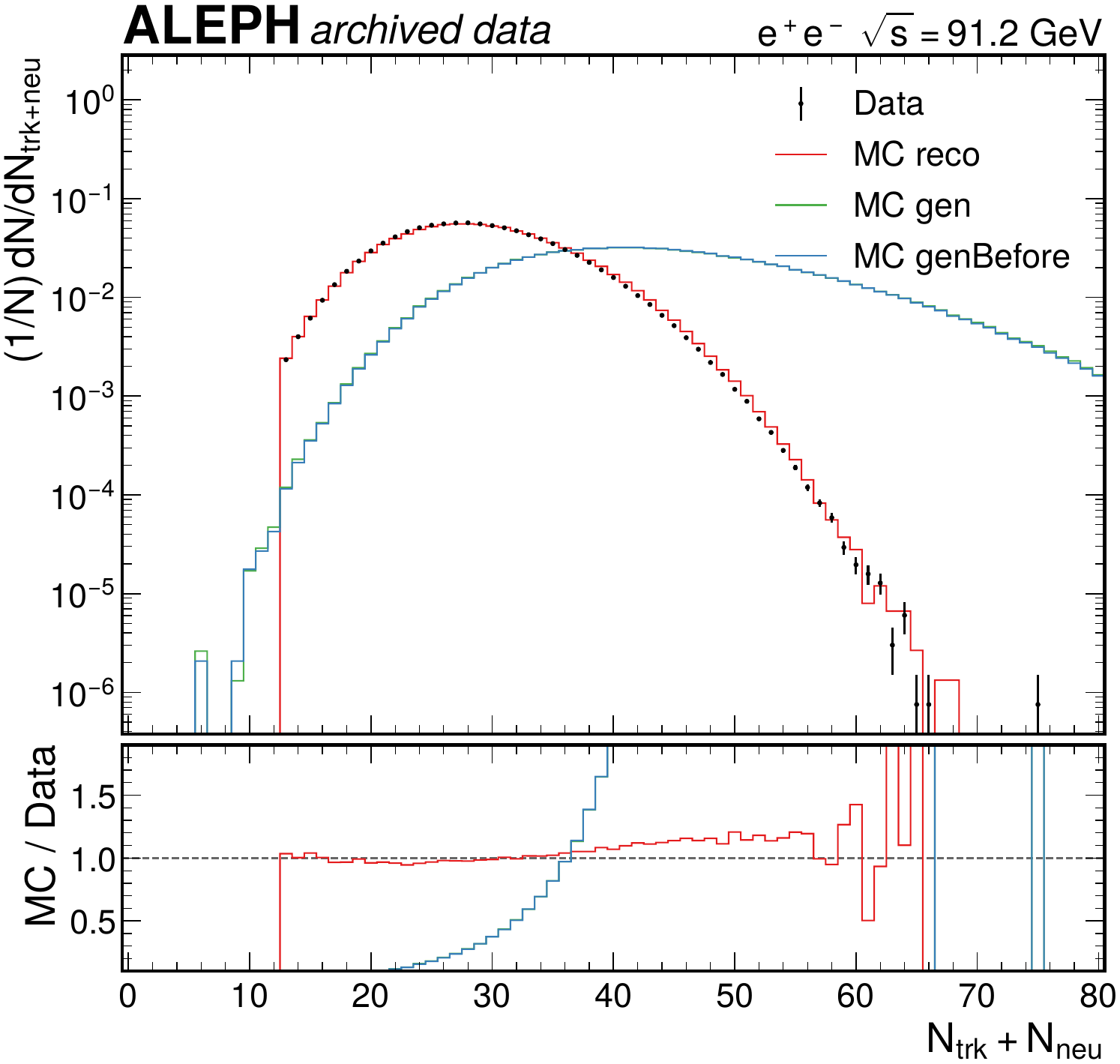}
\vspace{0.5em}
\sfpanel[0.495]{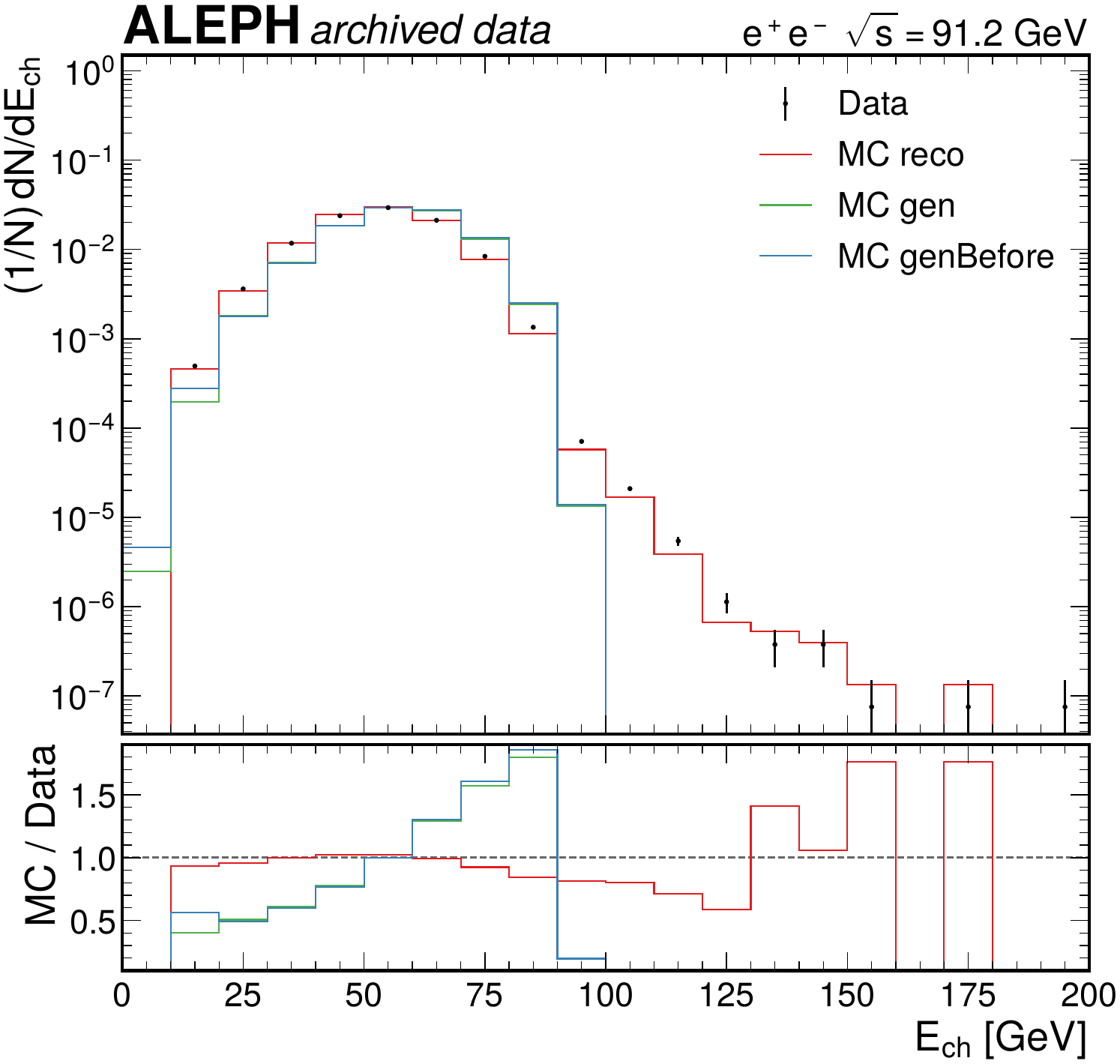}\hfill
\sfpanel[0.495]{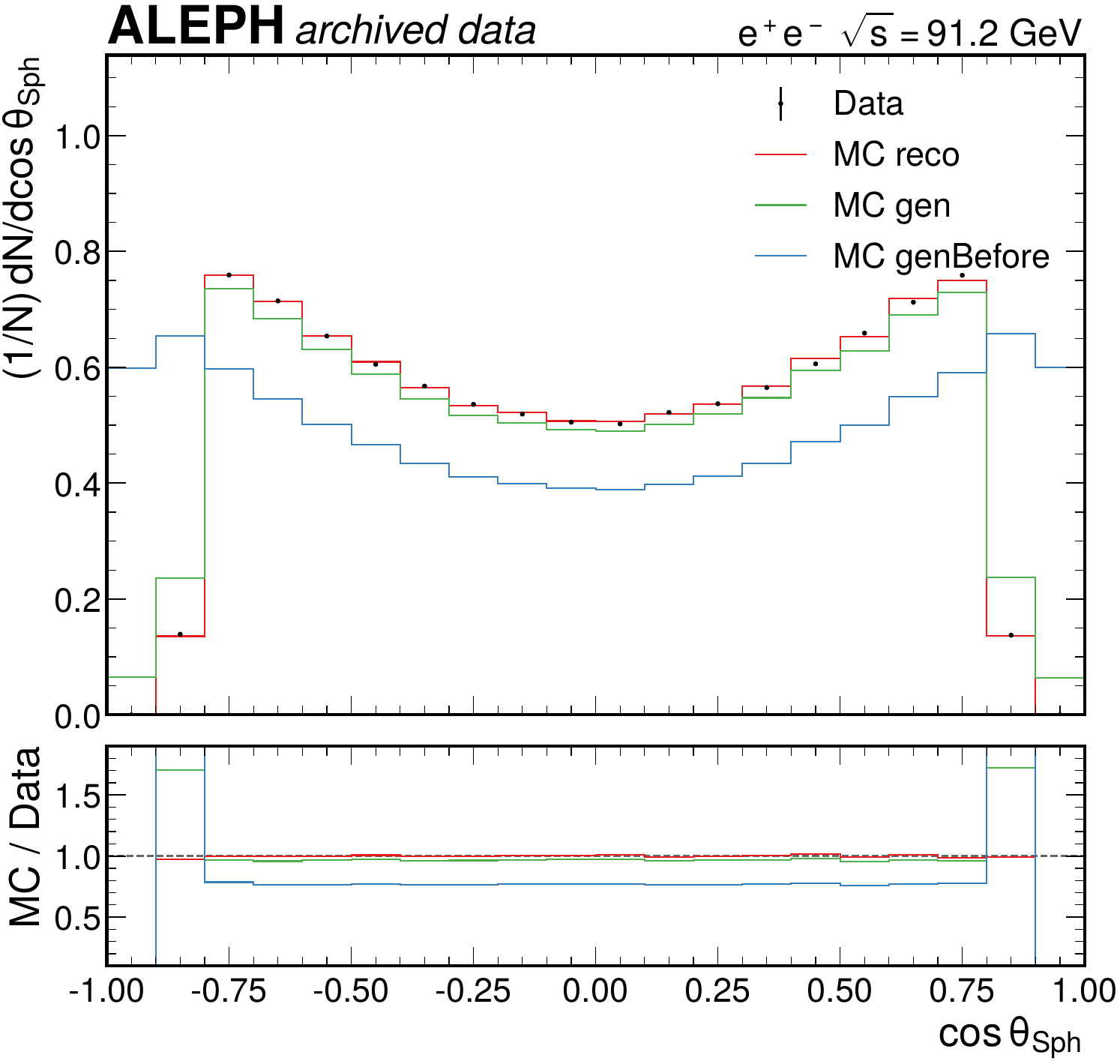}
\caption{Event-level observables used in the nominal hadronic selection; legends include detector-level data/MC and generator-level references. The (a) number of charged tracks, (b) number of charged tracks plus neutral particles, (c) total charged energy, and (d) cosine of the sphericity axis polar angle are shown. The MC\,Gen and MC\,GenBefore ratios display an even-odd staggering: generator-level charged multiplicity has an approximately 6:1 even-to-odd ratio (arising from the Z-decay net-charge constraint and the specific ALEPH Pythia~6 particle record), while reconstruction efficiency losses smear individual tracks one at a time and equalise the even/odd balance in data and MC\,Reco. MC\,Reco is the relevant comparison for the analysis.}
\label{fig:event_selection_observables}
\end{figure}

The agreement between the data and MC samples is assessed below. The particle and track level distributions are shown after all selections. The energy-flow candidates are categorized as charged tracks (Figure~\ref{fig:pwflag0_full}), charged lepton 1 (electron) candidates (Figure~\ref{fig:pwflag1_full}), charged lepton 2 (muon) candidates (Figure~\ref{fig:pwflag2_full}), neutral long-lived particles that decay to two charged tracks ($V^{0}$) (Figure~\ref{fig:pwflag3_full}), photons (Figure~\ref{fig:pwflag4_full}), and neutral hadrons (Figure~\ref{fig:pwflag5_full}) based on the reconstructed properties utilizing different detector subsystems. Event-level variables used for hadronic selection are shown in Figure~\ref{fig:event_selection_observables}. The selections on these observables define the phase space used to build detector response and unfolded event-shape spectra. The data-MC agreement is in general quite good with some mismodeling observed in particular for the neutral particles.

Several effects are visible in these distributions. An enhancement near $\cos\theta=0$
appears in both data and simulation for charged tracks, as shown in Figure~\ref{fig:pwflag0_cosTheta}. It originates from the fixed floating-point precision used when the archived sample was reprocessed and will be addressed in a subsequent reprocessing. Narrow peaks at $E\simeq|\vec{p}|\simeq45$~GeV appear in the charged-track (Figures~\ref{fig:pwflag0_pt},~\ref{fig:pwflag0_pmag},~\ref{fig:pwflag0_energy}) and $V^{0}$ (Figures~\ref{fig:pwflag3_pt},~\ref{fig:pwflag3_pmag},~\ref{fig:pwflag3_energy}) spectra of the reconstructed simulation but not at generator level. These are reconstruction-level effects, and their origins remain under study. The muon $\phi$ spectrum in Figure~\ref{fig:pwflag2_phi} shows an enhancement present only at generator level, which is believed to arise from the processing of the archived simulation. The $\mathrm{N}_\mathrm{trk}$ oscillatory structure for the generator-level distributions in Figure~\ref{fig:ntrk} is an artifact of the following: charged particles are produced in pairs, so even bins are populated, while radiation effects populate odd bins. The impact of these effects on the thrust spectrum is negligible.


\begin{figure}[p]
\centering
\vspace*{\fill}
\pwpanel{pwflag0}{cosTheta}\hfill
\pwpanel{pwflag0}{phi}\hfill
\pwpanel{pwflag0}{pt}
\vfill
\pwpanel{pwflag0}{pmag}\hfill
\pwpanel{pwflag0}{d0}\hfill
\pwpanel{pwflag0}{z0}
\vfill
\pwpanel{pwflag0}{ntpc}\hfill
\pwpanel{pwflag0}{energy}\hfill
\pwpanel{pwflag0}{mass}
\vspace*{\fill}
\caption{Selection variables for charged tracks: (a) $\cos\theta$, (b) $\phi$, (c) transverse momentum \pT, (d) momentum magnitude $|\vec{p}|$, (e) transverse impact parameter $d_{0}$, (f) longitudinal impact parameter $z_{0}$, (g) TPC hit multiplicity $N_{\mathrm{TPC}}$, (h) energy, and (i) mass.}
\label{fig:pwflag0_full}
\end{figure}

\begin{figure}[p]
\centering
\vspace*{\fill}
\pwpanel{pwflag1}{cosTheta}\hfill
\pwpanel{pwflag1}{phi}\hfill
\pwpanel{pwflag1}{pt}
\vfill
\pwpanel{pwflag1}{pmag}\hfill
\pwpanel{pwflag1}{d0}\hfill
\pwpanel{pwflag1}{z0}
\vfill
\pwpanel{pwflag1}{ntpc}\hfill
\pwpanel{pwflag1}{energy}\hfill
\pwpanel{pwflag1}{mass}
\vspace*{\fill}
\caption{Selection variables for electrons: (a) $\cos\theta$, (b) $\phi$, (c) transverse momentum \pT, (d) momentum magnitude $|\vec{p}|$, (e) transverse impact parameter $d_{0}$, (f) longitudinal impact parameter $z_{0}$, (g) TPC hit multiplicity $N_{\mathrm{TPC}}$, (h) energy, and (i) mass.}
\label{fig:pwflag1_full}
\end{figure}

\begin{figure}[p]
\centering
\vspace*{\fill}
\pwpanel{pwflag2}{cosTheta}\hfill
\pwpanel{pwflag2}{phi}\hfill
\pwpanel{pwflag2}{pt}
\vfill
\pwpanel{pwflag2}{pmag}\hfill
\pwpanel{pwflag2}{d0}\hfill
\pwpanel{pwflag2}{z0}
\vfill
\pwpanel{pwflag2}{ntpc}\hfill
\pwpanel{pwflag2}{energy}\hfill
\pwpanel{pwflag2}{mass}
\vspace*{\fill}
\caption{Kinematics for muons: (a) $\cos\theta$, (b) $\phi$, (c) transverse momentum \pT, (d) momentum magnitude $|\vec{p}|$, (e) transverse impact parameter $d_{0}$, (f) longitudinal impact parameter $z_{0}$, (g) TPC hit multiplicity $N_{\mathrm{TPC}}$, (h) energy, and (i) mass.}
\label{fig:pwflag2_full}
\end{figure}


\begin{figure}[p]
\centering
\vspace*{\fill}
\pwpanel{pwflag3}{cosTheta}\hfill
\pwpanel{pwflag3}{phi}\hfill
\pwpanel{pwflag3}{pt}
\vfill
\pwpanel{pwflag3}{pmag}\hfill
\pwpanel{pwflag3}{d0}\hfill
\pwpanel{pwflag3}{z0}
\vfill
\pwpanel{pwflag3}{ntpc}\hfill
\pwpanel{pwflag3}{energy}\hfill
\pwpanel{pwflag3}{mass}
\vspace*{\fill}
\caption{Kinematics for $V^{0}$ candidates: (a) $\cos\theta$, (b) $\phi$, (c) $p_{\mathrm{T}}$, (d) $|\vec{p}|$, (e) $d_{0}$, (f) $z_{0}$, (g) $N_{\mathrm{TPC}}$, (h) energy, and (i) mass.}
\label{fig:pwflag3_full}
\end{figure}


\begin{figure}[t!]
\centering
\pwpanel[0.495]{pwflag4}{cosTheta}\hfill
\pwpanel[0.495]{pwflag4}{phi}
\vspace{0.5em}
\pwpanel[0.495]{pwflag4}{pmag}\hfill
\pwpanel[0.495]{pwflag4}{energy}
\caption{Kinematics for photons: (a) $\cos\theta$, (b) $\phi$, (c) $|\vec{p}|$, and (d) energy.}
\label{fig:pwflag4_full}
\end{figure}


\begin{figure}[t!]
\centering
\pwpanel[0.495]{pwflag5}{cosTheta}\hfill
\pwpanel[0.495]{pwflag5}{phi}
\vspace{0.5em}
\pwpanel[0.495]{pwflag5}{energy}\hfill
\pwpanel[0.495]{pwflag5}{mass}
\caption{Kinematics of neutral hadrons: (a) $\cos\theta$, (b) $\phi$, (c) energy, and (d) mass.}
\label{fig:pwflag5_full}
\end{figure}


\FloatBarrier

\section{Analysis Method}
\label{sec:method}

\subsection{Observable Definition and Construction}

\begin{figure}[ht!]
\centering
\includegraphics[width=0.85\textwidth]{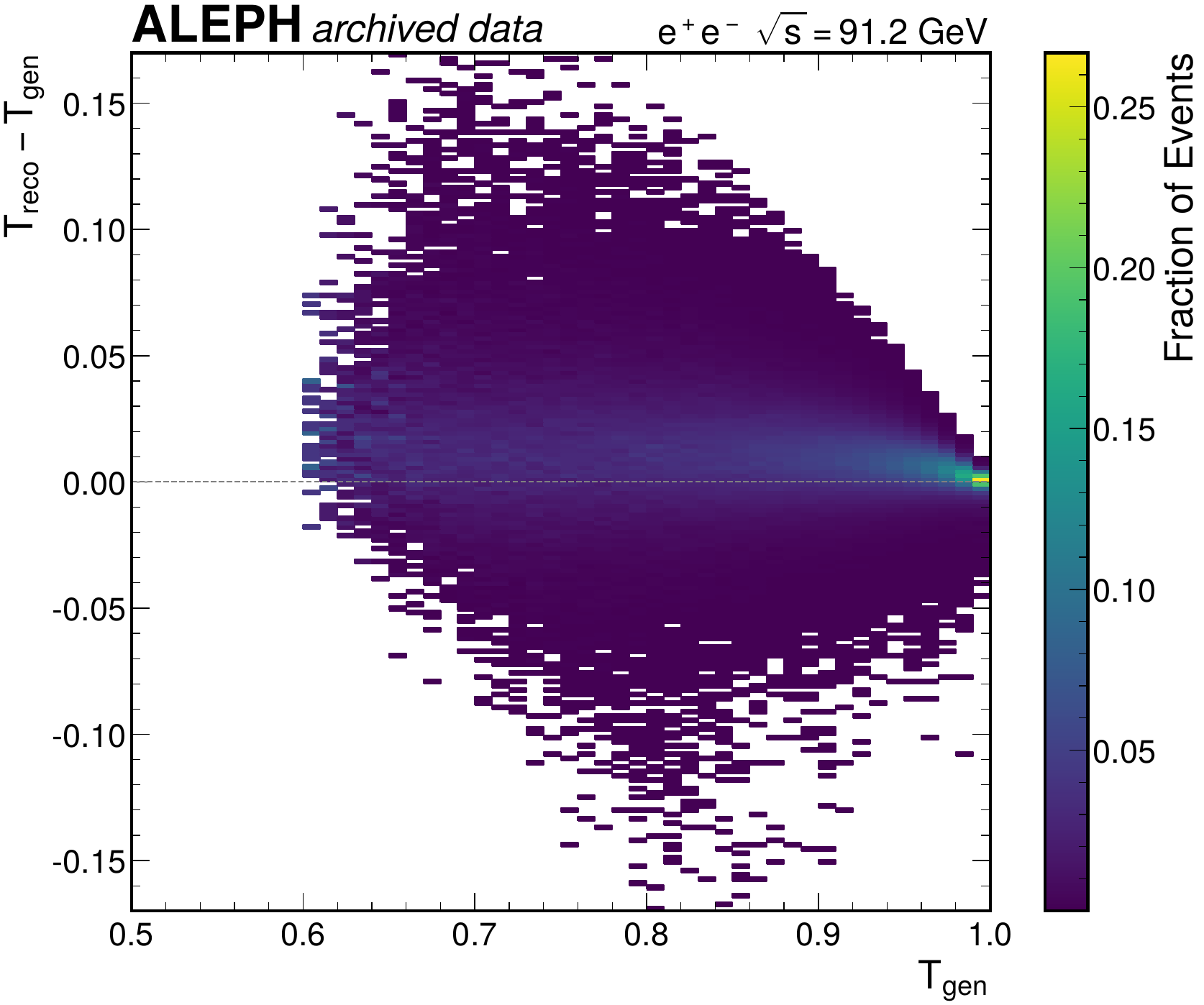}
\caption{Thrust resolution $(T_{\mathrm{reco}}-T_{\mathrm{gen}})$ versus $T_{\mathrm{gen}}$ for matched MC event pairs.}
\label{fig:thrust_resolution_2d}
\end{figure}

The thrust observable is defined as~\cite{PhysRevLett.39.1587}
\begin{equation}
T = \max_{\hat{n}} \frac{\sum_i |\vec{p}_i \cdot \hat{n}|}{\sum_i |\vec{p}_i|},
\end{equation}
with all selected charged and neutral reconstructed objects from Section~\ref{sec:event_selection} entering the sum. The thrust is built using selected charged and neutral particles. Those objects are input to an optimization algorithm that iteratively determines the thrust axis. The algorithm we use is a clone of the \textsc{Herwig++} implementation~\cite{Brandt:1978zm, B_hr_2008, BUCKLEY20132803, rivet_thrust}, which is an exact calculation rather than one based on heuristics as is done by others~\cite{Belle2Thrust}. The resolution is characterized using matched reco--gen event pairs. Figure~\ref{fig:thrust_resolution_2d} shows $(T_{\mathrm{reco}} - T_{\mathrm{gen}})$ as a function of $T_{\mathrm{gen}}$. The resolution narrows toward $T\to1$. To enable direct comparison, our work adopts the same binning as the previous ALEPH analyses~\cite{Barate:1996fi, Heister:2003aj}, with 42 uniform bins over $0.58 \le T < 1.0$ (bin width $\Delta T = 0.01$).

\subsection{Unfolding Setup and Regularization Choice}

\begin{figure}[ht!]
\centering
\includegraphics[width=0.85\textwidth]{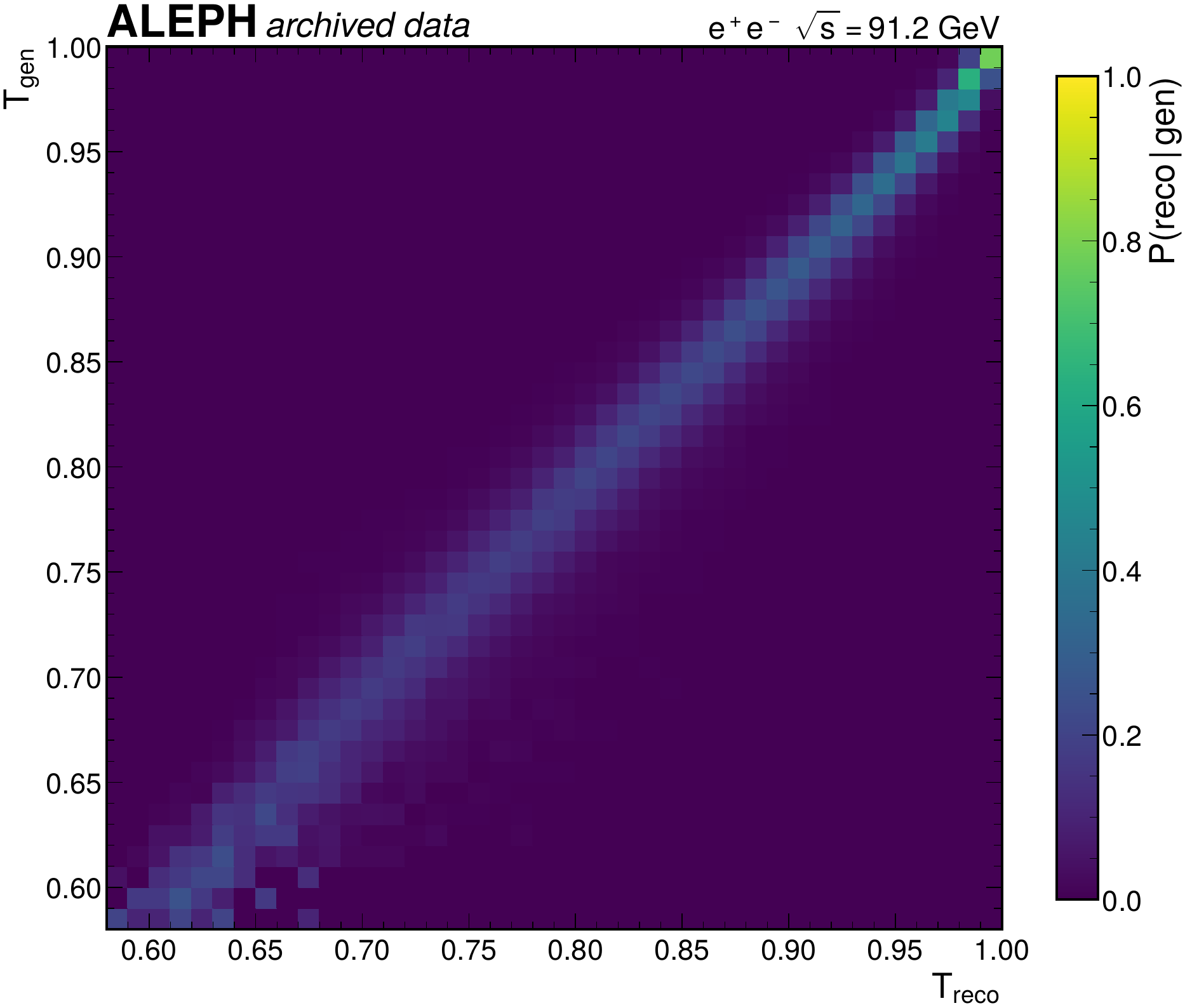}
\caption{Nominal response probability matrix for thrust.}
\label{fig:response_matrices_nominal}
\end{figure}

Detector correction is performed with the IBU method of D'Agostini~\cite{DAgostini:1994zf,dagostini2010improvediterativebayesianunfolding}, as implemented in RooUnfold~\cite{Adye:2011gm},
\begin{equation}
U_j^{(n+1)} = \frac{1}{\epsilon_j}\sum_i M_i\,\frac{R_{ij}\,\pi_j^{(n)}}{\sum_k R_{ik}\,\pi_k^{(n)}},
\end{equation}
where $M_i$ is the measured detector-level spectrum, $R_{ij}$ is the response, and $\pi_j^{(n)}$ is the iteration prior. The prior for the first iteration is taken as the generator-level MC truth distribution $\pi_j^{(0)} = N_{\mathrm{gen}}(j)$, the standard RooUnfold default. This choice initializes the unfolding close to the expected truth shape and avoids any flat-prior amplification of high-frequency fluctuations in the first iteration. The response matrix is filled with MC events that pass the detector-level selection and the corresponding generator-level event as shown in Figure~\ref{fig:response_matrices_nominal}. A selected event whose reconstructed thrust falls outside the unfolding range enters instead through RooUnfold's miss term, which supplies the efficiency factor $1/\epsilon_j$ appearing below.

The IBU regularization term for the number of iterations is kept at a nominal value of $N_{\mathrm{iter}}=5$ as found by previous studies~\cite{Badea:2019vey}, with $N_{\mathrm{iter}}=4$ and $6$ propagated as unfolding-regularization systematic branches.
The iteration-choice diagnostics, including refolded spectra at $N=4,5,6,10,100$ compared with detector-level data, confirm that the closure-shape $\chi^2/\mathrm{ndf}$ is stable for $N_{\mathrm{iter}}\in\{4,5,6\}$. The spread in the unfolded thrust spectrum across this range is below $1\%$ in 37 of the 42 reported bins (median $0.3\%$), rising to at most $6.6\%$ near the two-jet edge ($T\lesssim0.62$) where the steeply falling spectrum amplifies regularization sensitivity.

The response matrix has a diagonal fraction of $f_{\mathrm{diag}}=0.33$, meaning approximately two-thirds of events migrate out of their generated thrust bin at reconstruction level. This level of bin-to-bin migration is expected for thrust at ALEPH given the finite charged-particle momentum resolution and the sensitivity of thrust to soft particles near the event boundary. It implies that the unfolding correction is non-trivial and that the choice of regularization strength meaningfully affects the result. This also motivates the need to measure the final bin-by-bin correlation to assess the overall statistical strength.

\begin{figure}[ht!]
\centering
\includegraphics[width=0.49\textwidth]{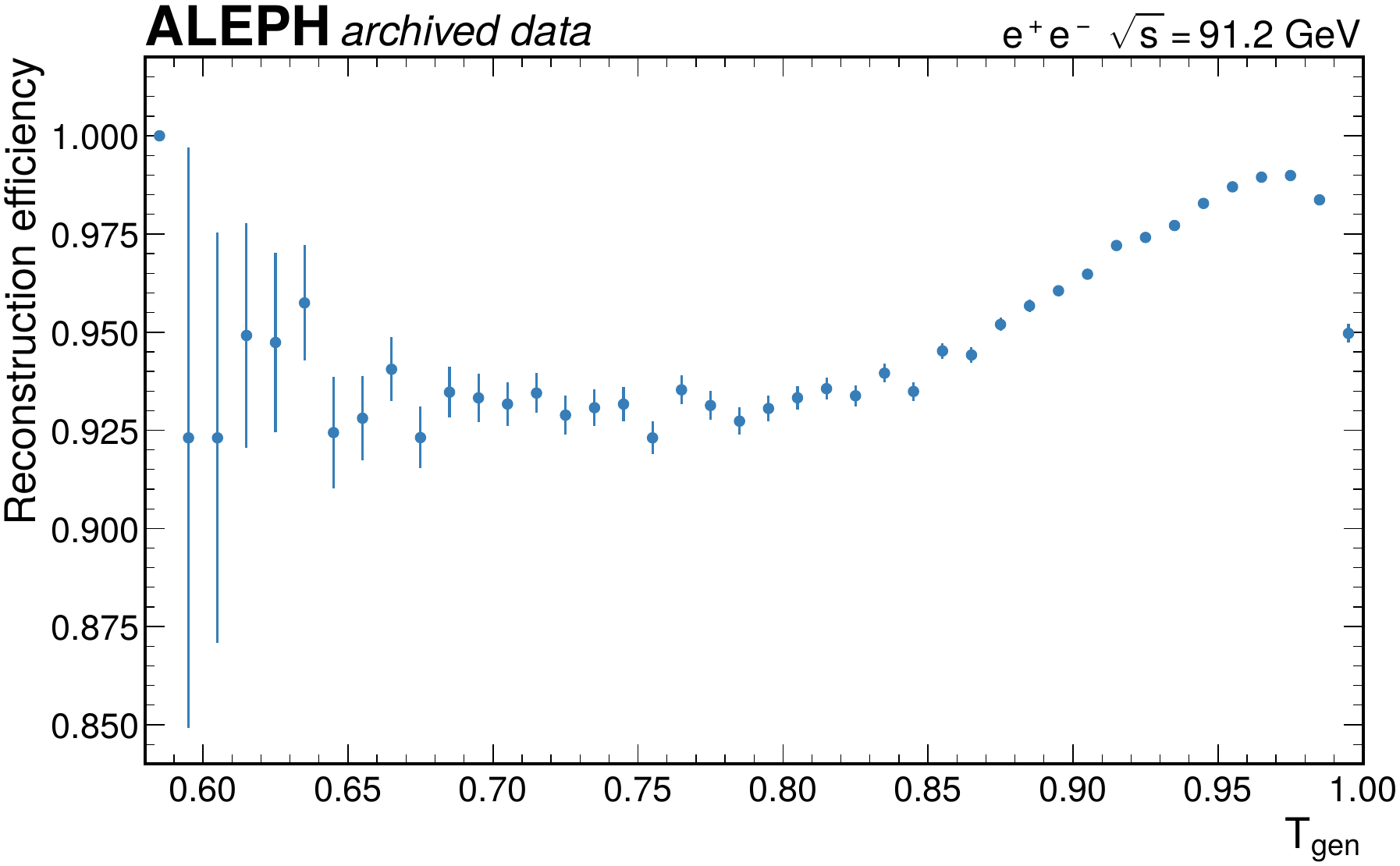}
\includegraphics[width=0.49\textwidth]{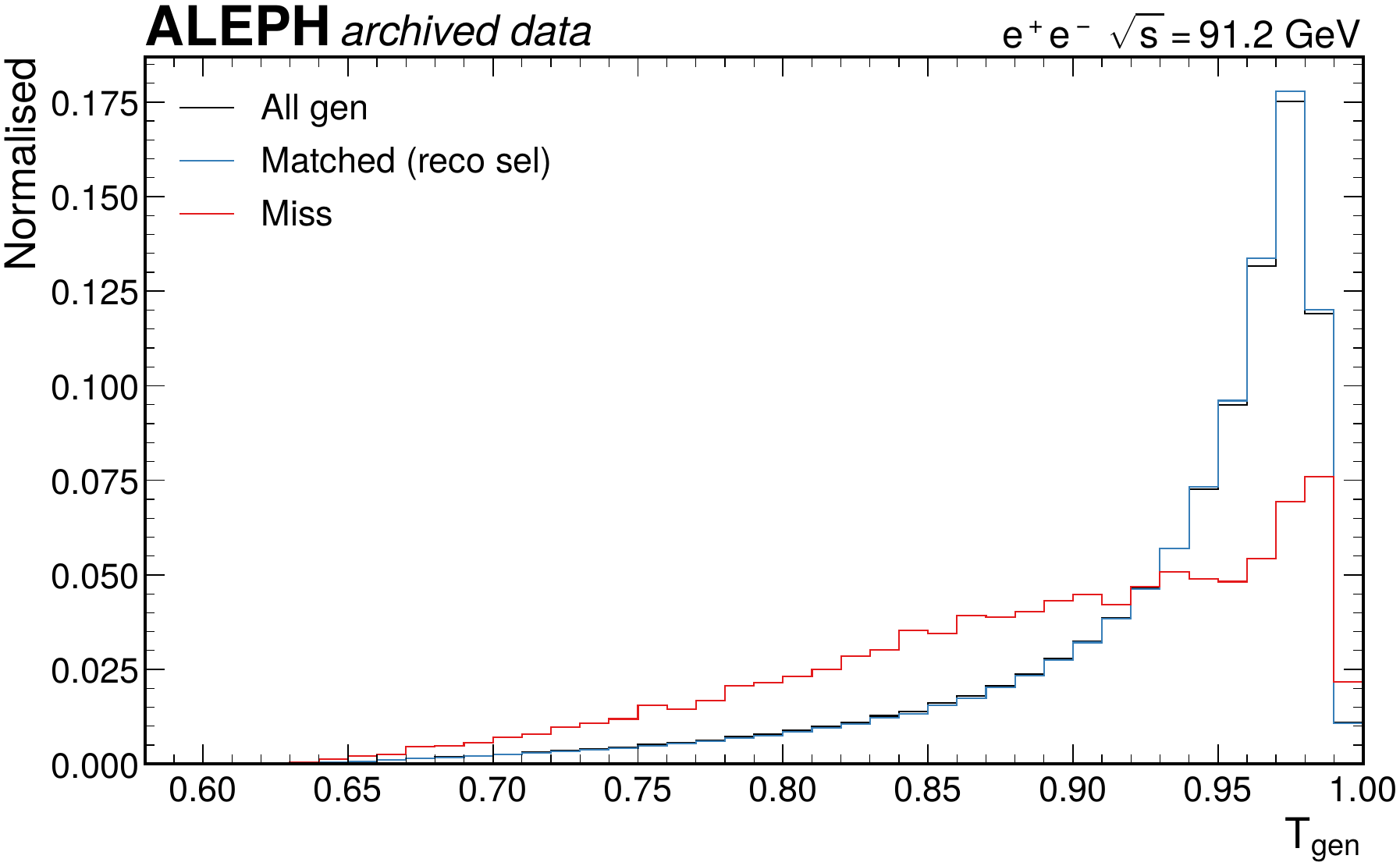}
\caption{Left: reco efficiency (matched/all gen) as a function of $T_{\mathrm{gen}}$,
with binomial uncertainties. Right: thrust distributions for all gen, matched, and miss
events (normalized to unit area). Miss events are concentrated at low thrust and low
multiplicity, consistent with events failing the hadronic event selection.}
\label{fig:miss_term_efficiency}
\end{figure}

The miss term, the generator-level events with no matched reconstructed counterpart, is
shown in Figure~\ref{fig:miss_term_efficiency}. These are events that fail the reco-level hadronic event selection despite being generated within the fiducial volume. The miss rate is concentrated at low thrust ($T\lesssim0.7$) and in events with low charged multiplicity, consistent with the hadronic event selection removing poorly reconstructed events. RooUnfold accounts for miss events via \texttt{response.Miss($x_{\mathrm{gen}}$)}, which corrects the unfolded spectrum for this acceptance effect.

\subsection{Hadronic and QED ISR Correction}
\label{sec:corrections_and_cov}

After unfolding, two multiplicative corrections are applied. A correction for the hadronic event-selection,
\begin{equation}
C_{\mathrm{had}}(j) = \frac{N_{\mathrm{genBefore}}(j)}{N_{\mathrm{gen}}(j)},
\end{equation}
accounts for the efficiency of the selection at the generator level. The correction factor is derived from the archived \textsc{Pythia}~6 sample~\cite{Sjostrand:2000wi} and is applied so that the final spectrum restores the yield lost by the hadronic selection.

Then a correction for electromagnetic (QED) initial-state radiation (ISR) is applied,
\begin{equation}
C_{\mathrm{ISR}}(j) = \frac{N^{\mathrm{gen}}_{\mathrm{no\,ISR}}(j)}{N^{\mathrm{gen}}_{\mathrm{ISR}}(j)}.
\end{equation}
This factor is applied to enable comparisons with state-of-the-art theory calculations which do not include QED ISR. The correction is derived from standalone \textsc{Pythia}~8.317 samples generated with ISR switched off and on~\cite{Bierlich:2022pfr}. \textsc{Pythia}~8.317 is chosen here for simplicity; a higher-order generator such as \textsc{KKMCee}~5.00.02 would be preferred for a full thrust measurement~\cite{Jadach:2022mbe}. The ratio histograms are pre-computed once into a cache file and applied identically to the nominal and all systematic variation branches, so that the ISR correction cancels in the systematic shifts and its residual uncertainty (from finite MC statistics on the ratio) is propagated as a separate component $\Delta_{\mathrm{ISR}}$ (Section~\ref{sec:isr_uncertainty}).

For the nominal result, the data statistical covariance is computed by RooUnfold using analytical error propagation (\texttt{RooUnfold::kCovariance}), which propagates Poisson uncertainties in the measured data through the IBU iteration chain without toy MC sampling. The response matrix's own finite statistics are not included by that call and are propagated separately by a Poisson bootstrap over the simulated events, as described in Section~\ref{sec:stat_uncertainty}. The resulting covariance matrices are read from the unfolding output and transformed to the normalized differential-thrust basis used in this work. Non-statistical component covariances are built with a configurable model, and all covariance products are exported over the reported bins.

\section{Systematic Uncertainties}
\label{sec:systematics}

Systematic uncertainties arise from several components: data statistical, MC statistical, unfolding-regularization, experimental variations, theory unfolding prior, hadronic event-selection correction, and ISR correction. The sections below describe each component.

\subsection{Statistical Uncertainty}
\label{sec:stat_uncertainty}

The data statistical uncertainty is propagated analytically by RooUnfold through the IBU iteration chain, without toy-MC sampling, yielding a full statistical covariance matrix $\mathrm{Cov}^{(\mathrm{stat})}_{ij}$. The per-bin fractional uncertainty $\Delta_{\mathrm{stat}}(j)=\sqrt{\mathrm{Cov}^{(\mathrm{stat})}_{jj}}/U_j^{\mathrm{corr}}$ is typically below $2\%$ in the central range and nearly uniform across the spectrum. The uncertainty is shown in Figure~\ref{fig:unc_comp_stat_iter}.

MC statistical uncertainties enter the budget through the two corrections derived from finite simulated samples: the MC statistical uncertainty of the ISR-off and ISR-on samples is propagated onto $C_{\mathrm{ISR}}(j)$, and the archived-MC statistical uncertainty of $N_{\mathrm{genBefore}}$ and $N_{\mathrm{gen}}$ is propagated onto the hadronic event-selection correction $C_{\mathrm{had}}(j)$. The finite statistics of the response matrix itself are propagated separately, by giving every matched simulated event a Poisson weight and re-unfolding the data over 200 such replicas; the sample covariance of the resulting spectra is taken as $\mathrm{Cov}^{(\mathrm{MCstat})}_{ij}$. This is required because \texttt{RooUnfold::kCovariance} transports the measured data covariance only, and because the simulated sample is smaller than the data sample, so the response statistical term exceeds the data statistical term in every bin.

\subsection{Unfolding Regularization Uncertainty}
\label{sec:regularization_uncertainty}

\begin{figure}[ht!]
\centering
\includegraphics[width=0.49\textwidth]{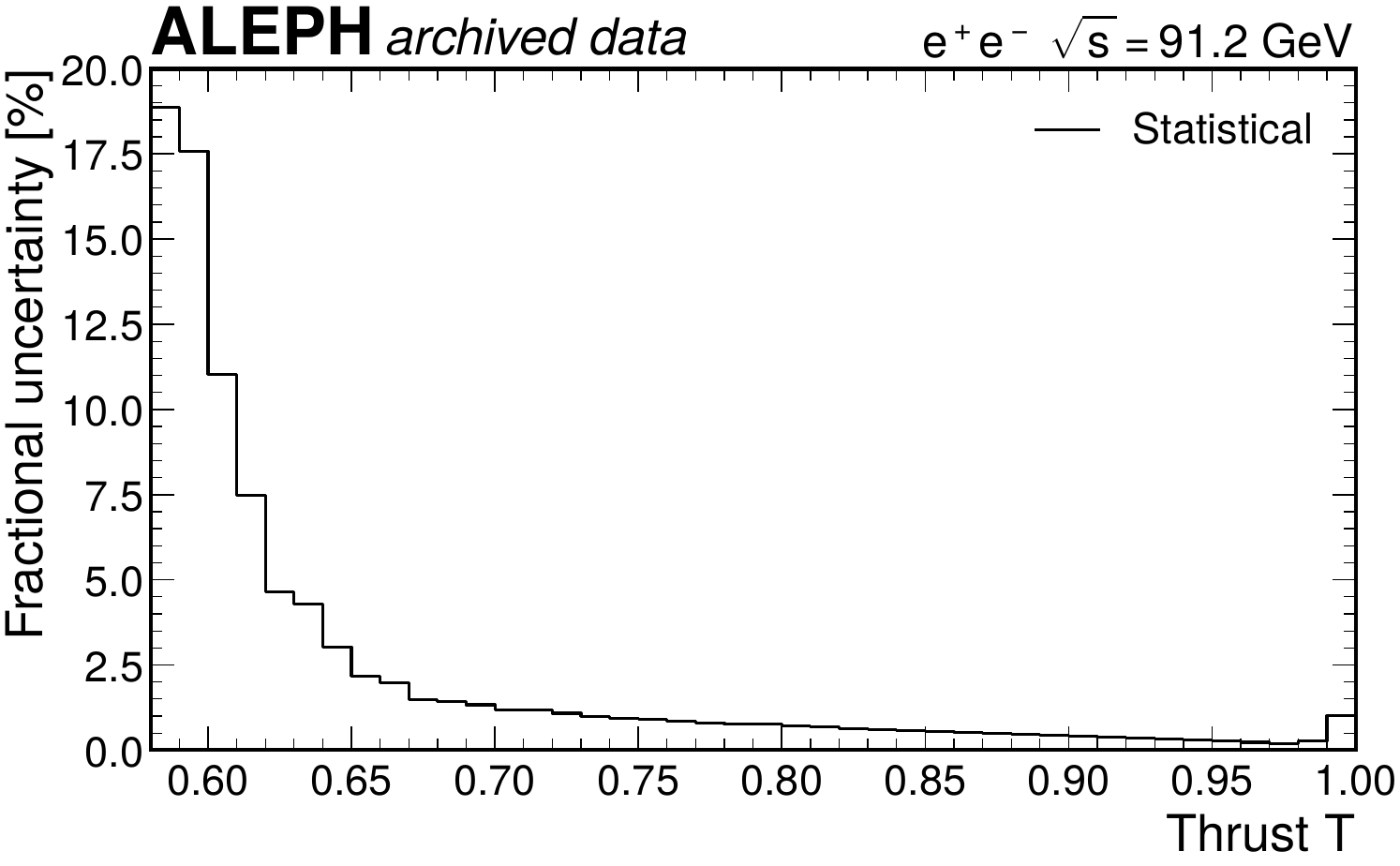}%
\hfill%
\includegraphics[width=0.49\textwidth]{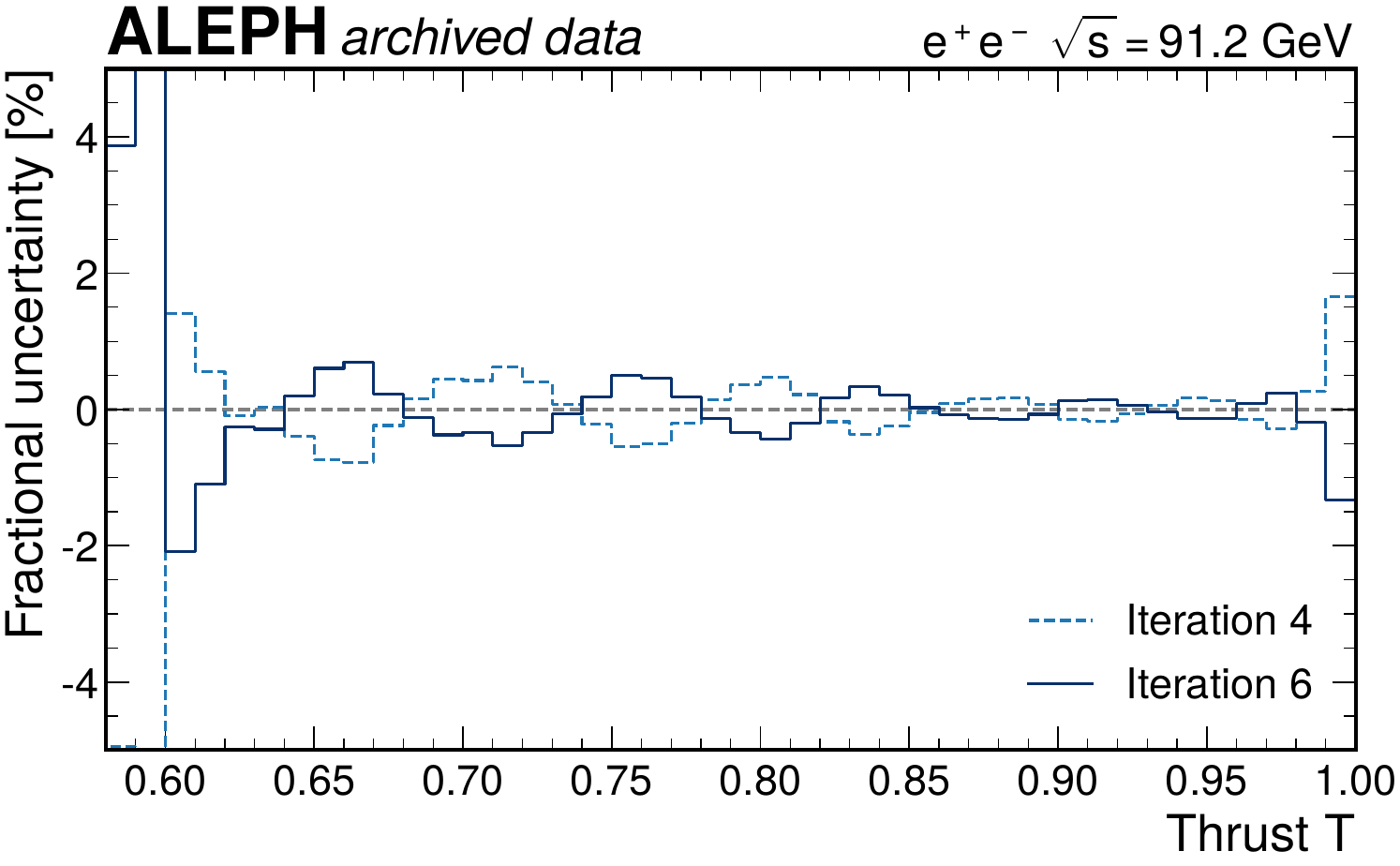}
\caption{Signed per-bin shifts for the statistical (left) and unfolding-iteration
(right) uncertainty components. The statistical component is nearly flat and small
across the spectrum. The iteration panel shows $N_{\mathrm{iter}}=4$ and $6$
relative to the nominal $N_{\mathrm{iter}}=5$.}
\label{fig:unc_comp_stat_iter}
\end{figure}

The unfolding regularization uncertainty is closely related: two additional unfolding passes at $N_{\mathrm{iter}}=4$ and $N_{\mathrm{iter}}=6$ quantify the sensitivity of the result to the regularization choice. The iteration envelope $\Delta_{\mathrm{iter}}(j)=\max(|\delta_{\mathrm{iter4}}(j)|,|\delta_{\mathrm{iter6}}(j)|)$
is sub-percent in the bulk of the spectrum, confirming that the nominal $N_{\mathrm{iter}}=5$ sits on a stable plateau and that regularization is not a limiting source of uncertainty. Figure~\ref{fig:unc_comp_stat_iter} shows the
signed per-bin shifts for both components across the measurement range. The iteration shifts change sign across the range because $N_{\mathrm{iter}}=4$ and $6$ bracket the nominal, and the bin-to-bin alternation at low thrust reflects the anti-correlation that regularization induces between neighboring bins.

\FloatBarrier

\subsection{Experimental Systematic Uncertainties}
\label{sec:exp_systematics}

Experimental systematic uncertainties are handled following the standard ALEPH procedure~\cite{Heister:2003aj} to vary individual selections, rerunning the full analysis chain, and measuring the difference with respect to the nominal as the uncertainty. The explored systematics are defined below.

\begin{itemize}
  \item \texttt{iter4}/\texttt{iter6}: unfolding passes at $N_{\mathrm{iter}}=4$ and $6$.
  \item \texttt{ntpc7}: tightened TPC hit requirement ($N_{\mathrm{TPC}}>7$, nominal $>4$).
  \item \texttt{pt04}: raised charged-$p_T$ threshold ($>0.4$~GeV, nominal $>0.2$~GeV).
  \item \texttt{ech10}: loosened charged-energy cut ($>10$~GeV, nominal $>15$~GeV).
  \item \texttt{with\_met}: missing transverse energy added as a pseudo-particle.
  \item \texttt{nes\_up}/\texttt{nes\_down}: neutral energy scale varied up and down.
  \item \texttt{ner}: neutral objects randomly removed with 2.5\%, probing acceptance efficiency.
\end{itemize}

\begin{figure}[ht!]
\centering
\includegraphics[width=0.85\textwidth]{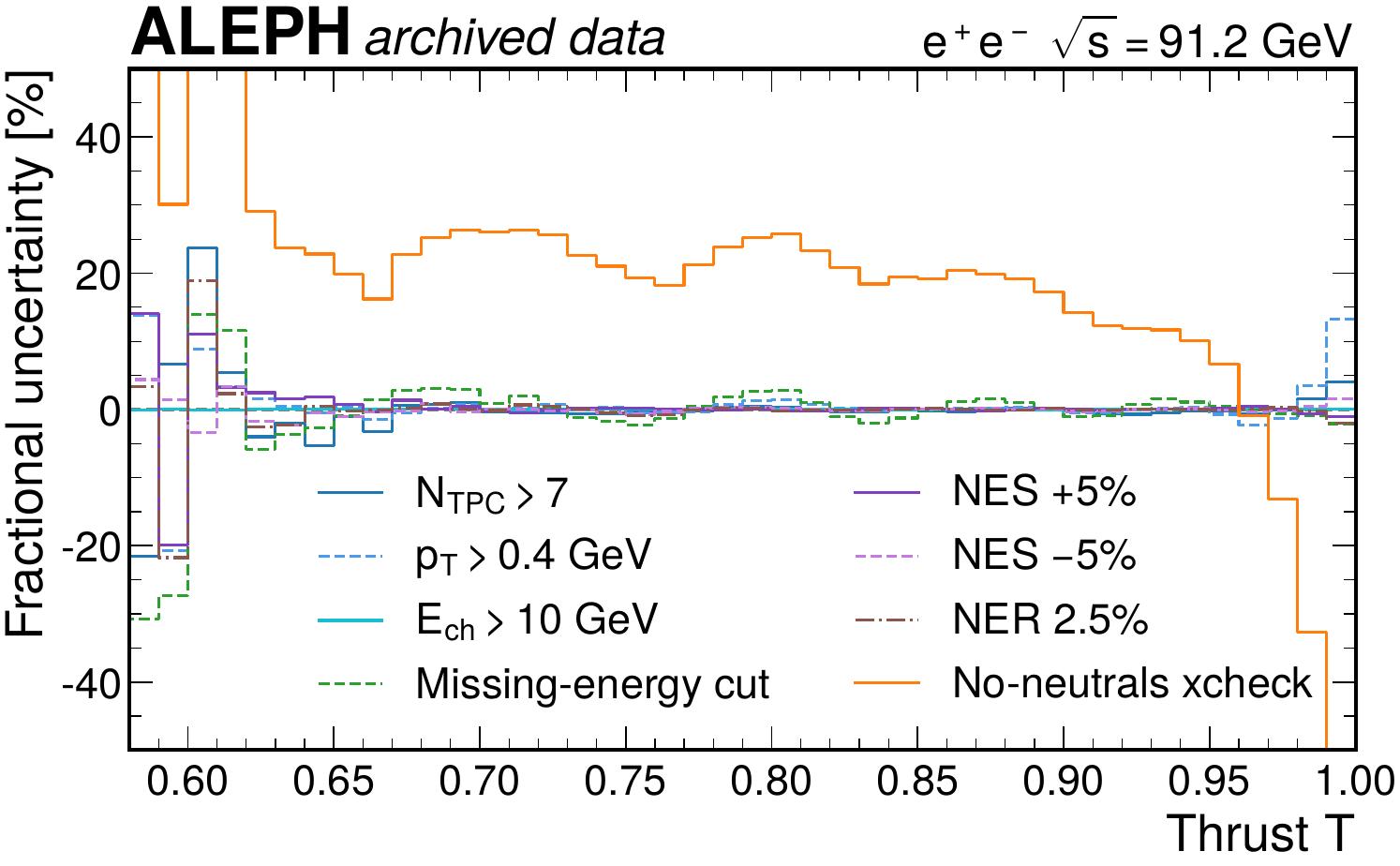}
\caption{Signed per-bin shifts for the seven experimental variation branches: the four
track/event branches (\texttt{ntpc7}, \texttt{pt04}, \texttt{ech10},
\texttt{with\_met}) and the three neutral-response branches (\texttt{nes\_up},
\texttt{nes\_down}, \texttt{ner}). The \texttt{no\_neutrals} curve is a cross-check and
does not enter the uncertainty.}
\label{fig:unc_comp_exp}
\end{figure}

Figure~\ref{fig:unc_comp_exp} shows the signed per-bin shifts for all variations. The largest contributions arise at low thrust ($T\lesssim 0.62$) where the steeply rising spectrum amplifies small absolute shifts; across the central range the experimental envelope is at the few-percent level.

The neutral-object response variations are chosen in preference to the standard ALEPH approach of removing all neutral particles within 1--2 GeV from the event~\cite{Heister:2003aj}. The reason is that
removing neutrals entirely redefines the observable: the resulting distribution is no longer the energy-flow thrust $T_{\mathrm{EF}}$ but a different quantity, charged-only thrust $T_{\mathrm{ch}}$, computed from a different particle-level definition. Treating the difference $T_{\mathrm{EF}} - T_{\mathrm{ch}}$ as a systematic uncertainty on the energy-flow measurement would conflate two distinct observables and produce an uncertainty that is unphysically large. Instead, \texttt{nes\_up}, \texttt{nes\_down}, and \texttt{ner} vary the detector response to neutral objects while holding the particle-level definition fixed, providing a physically meaningful probe of calorimeter modeling uncertainties without changing what is being measured.

\begin{figure}[t!]
\centering
\includegraphics[width=0.70\textwidth]{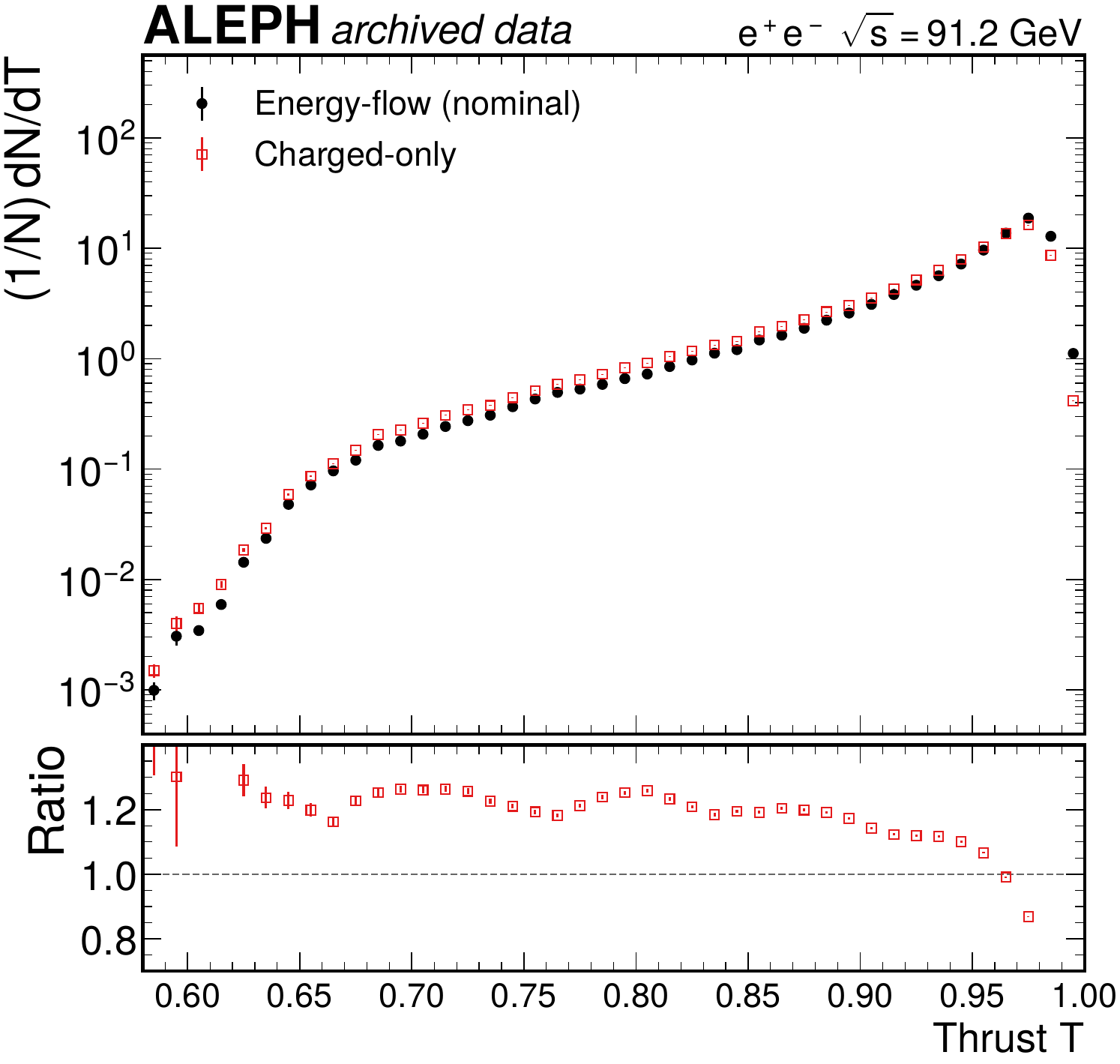}
\caption{Comparison of the charged-only (red open circles) and energy-flow (black filled circles) unfolded thrust distributions. Error bars represent statistical uncertainties only.}
\label{fig:charged_vs_ef}
\end{figure}

To quantify this distinction, the full unfolding chain was repeated with neutral particles ($\mathrm{pwflag}\in\{4,5\}$) removed, yielding a charged-only thrust distribution as shown in Figure~\ref{fig:charged_vs_ef}. Both the energy-flow and charged-only results are processed through the complete analysis chain. The reconstructed particle collection differs between the two, but both are unfolded to the same energy-flow particle-level definition. Both distributions are shown with statistical uncertainties only; the full systematic suite is not propagated for this diagnostic comparison, and the charged-only branch is excluded from the nominal total uncertainty combination. The charged-only spectrum is systematically broader than the energy-flow, as expected from the missing neutral-energy deposits.

The track and neutral object variations are combined as described below. The full experimental envelope is then combined in quadrature.

\begin{align}
\Delta_{\mathrm{track/event}}(j) &=
  \sqrt{\sum_{v\in\{\mathrm{ntpc7,\,pt04,\,ech10,\,with\_met}\}} |\delta_v(j)|^2}. \\
  \Delta_{\mathrm{neutral}}(j) & =
\sqrt{\max\!\left(|\delta_{\mathrm{nes\_up}}(j)|,|\delta_{\mathrm{nes\_down}}(j)|\right)^2
    + |\delta_{\mathrm{ner}}(j)|^2}. \\ 
\Delta_{\mathrm{exp}}(j) &= \sqrt{\Delta_{\mathrm{track/event}}^2(j)+\Delta_{\mathrm{neutral}}^2(j)}.
\end{align}

\FloatBarrier

\subsection{Theory Systematic Uncertainties}
\label{sec:theory_systematics}

\begin{figure}[ht!]
\centering
\includegraphics[width=0.85\textwidth]{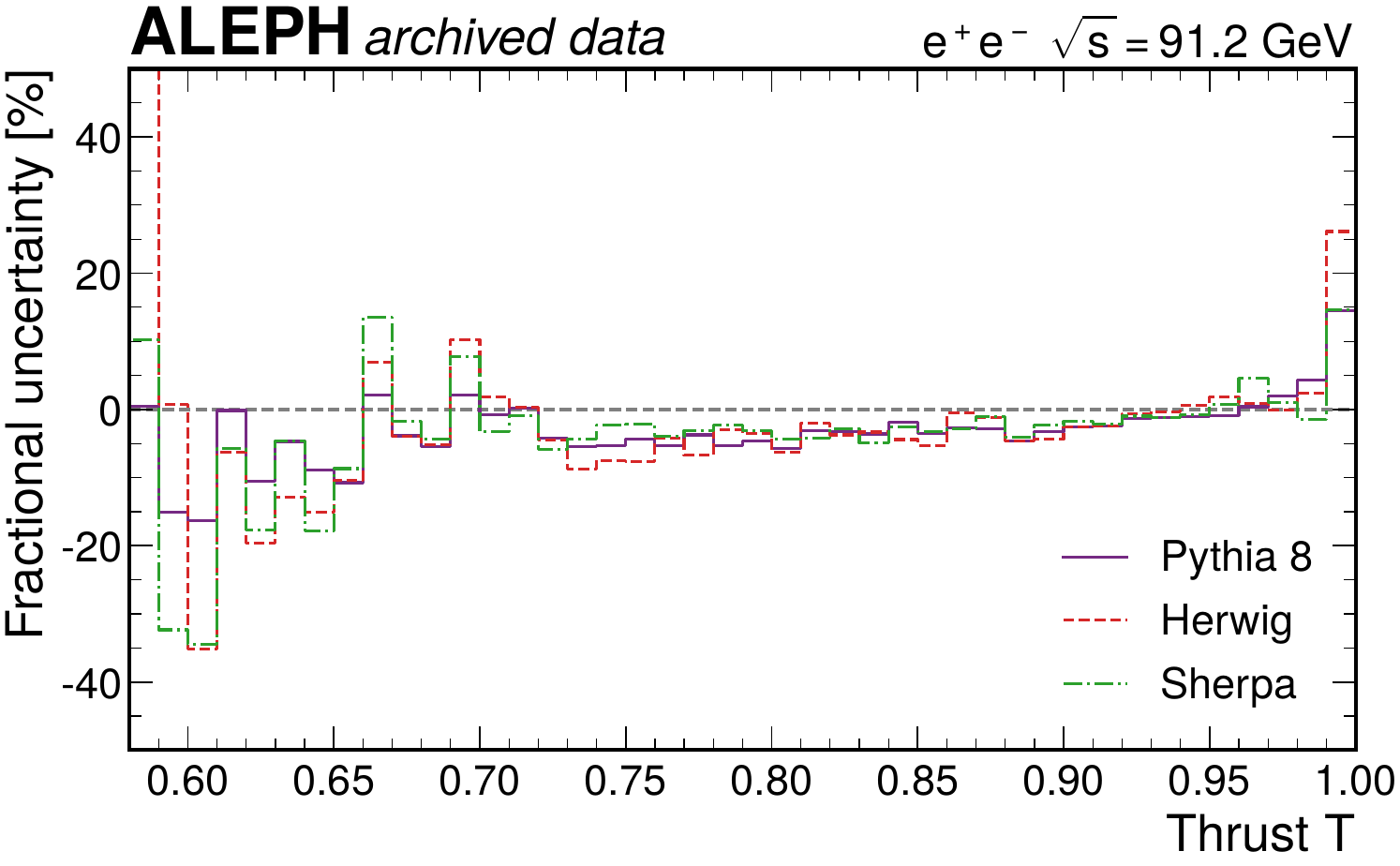}
\caption{Signed per-bin shifts for the theory prior variations from \textsc{Pythia}~8.317, \textsc{Herwig}~7.1.5, and \textsc{Sherpa}~2.2.6.}
\label{fig:unc_comp_theory}
\end{figure}

Theory systematic variations probe the residual model dependence introduced by the Monte Carlo generator used in the unfolding. The nominal response matrix is constructed from \textsc{Pythia}~6~\cite{Sjostrand:2000wi}, the ALEPH simulation. Three additional generators are used to reweight the generator-level thrust distribution event-by-event, producing alternative response matrices that span a
wide range of parton-shower algorithms and hadronization models: \textsc{Pythia}~8.317 with Monash tune, dipole-ordered shower~\cite{Bierlich:2022pfr}, \textsc{Herwig}~7.1.5, angular-ordered shower with cluster hadronization~\cite{B_hr_2008,Bellm:2015jjp,Bellm:2017bvx, bewick2024herwig73releasenote}, and \textsc{Sherpa}~2.2.6 with AHADIC++ hadronization~\cite{Bothmann2019}.
The theory envelope is taken as the maximum absolute shift across the three generators:
\begin{equation}
\Delta_{\mathrm{theory}}(j)=
  \max\!\left(|\delta_{\mathrm{pythia8}}(j)|,
             |\delta_{\mathrm{herwig}}(j)|,
             |\delta_{\mathrm{sherpa}}(j)|\right).
\end{equation}
The three generators diverge progressively at lower thrust, where multi-jet configurations and hard, wide-angle gluon emission dominate and the parton-shower and matrix-element matching differ between generators; there the reweighted
generator-level spectra move coherently and by similar amounts. The envelope also rises in the last bin, $0.99\le T<1.00$, where the reweighted spectra are least well constrained: the bin sits at the edge of the two-jet limit, and the shift there is not smoothly connected to its neighbours. The theory envelope is the dominant uncertainty contribution across the central measurement range, as shown in Figure~\ref{fig:unc_comp_theory}.

\FloatBarrier

\subsection{Hadronic Event-Selection Correction}
\label{sec:had_uncertainty}

After unfolding, a bin-wise hadronic event-selection correction is applied to account for the fraction of hadronic events lost to the generator-level selection:
\begin{equation}
C_{\mathrm{had}}(j) = \frac{N_{\mathrm{genBefore}}(j)}{N_{\mathrm{gen}}(j)},
\qquad
U_j^{\mathrm{corr}} = U_j\,C_{\mathrm{had}}(j).
\end{equation}
This correction maps the unfolded spectrum from the generator-level hadronic-event-selected phase space to the full inclusive hadronic phase space. The uncertainty on this correction is propagated as an independent component
$\Delta_{\mathrm{had}}$. The correction is computed from two generator-level histograms: \texttt{tgenBefore} (events before hadronic selection) and \texttt{tgen} (events passing hadronic selection). Statistical uncertainties on both histograms are propagated through the ratio using standard error propagation, yielding symmetric bin-wise up and down variations $C_{\mathrm{had}}^{\pm}(j)$. These variations are applied to the unfolded spectrum and normalized to the differential cross-section basis, producing up and down shifted spectra. 


\subsection{QED ISR Correction}
\label{sec:isr_uncertainty}

The QED ISR correction removes the contribution of events
where the $e^+e^-$ center-of-mass energy is reduced by photon radiation before the hard scatter, mapping the measured distribution from the ISR-inclusive sample
to the pure $\sqrt{s}=M_Z$ distribution:
\begin{equation}
C_{\mathrm{ISR}}(j) = \frac{N^{\mathrm{gen}}_{\mathrm{no\,ISR}}(j)}{N^{\mathrm{gen}}_{\mathrm{ISR}}(j)}.
\end{equation}
Both published ALEPH event-shape analyses apply this
correction~\cite{Barate:1996fi,Heister:2003aj}. The 1998 ALEPH analysis~\cite{Barate:1996fi} defines it as the ratio of generator-level thrust distributions produced with and without ISR, using \textsc{JETSET}~7.4. The 2004 analysis~\cite{Heister:2003aj} uses \textsc{KORALZ}, which provides a more accurate treatment of ISR, to generate the corrected distribution.

The ISR correction is applied bin-by-bin after the hadronic correction using standalone \textsc{Pythia}~8.317 samples with ISR disabled (\texttt{isr0}) and enabled (\texttt{isr1}), as described in Section~\ref{sec:corrections_and_cov}. The uncertainty on
this correction is derived from the finite Monte Carlo statistics. Up and down variations $C_{\mathrm{ISR}}^{\pm}(j) = C_{\mathrm{ISR}}(j)\cdot(1\pm\sigma_{\mathrm{rel}}(j))$ are applied to the unfolded spectrum and the envelope is taken as $\Delta_{\mathrm{ISR}}(j)$. The uncertainty is purely MC-statistical in nature for simplicity, but should be assessed in closer detail for future work. 


\subsection{Covariance Evaluation}
\label{sec:stat_covariance}

For the nominal run, the statistical covariance is computed by RooUnfold using analytical error propagation (\texttt{RooUnfold::kCovariance}), which propagates
Poisson uncertainties in the data and response histograms through the IBU iteration chain without toy MC sampling. The resulting covariance matrix is read from the unfolding output and transformed to the normalized differential-thrust basis used in this work.

Non-statistical component covariances are built with a configurable model. The current nominal release uses the \texttt{template\_correlated} option, where each non-statistical component covariance is built as
\begin{equation}
\mathrm{Cov}^{(c)}_{ij}=\Delta_c(i)\,\Delta_c(j)\,\rho^{\mathrm{stat}}_{ij},
\end{equation}
with $\rho^{\mathrm{stat}}_{ij}$ taken from the statistical unfolding covariance correlation template. This keeps component magnitudes tied to the measured per-bin
shifts while avoiding a purely diagonal systematic-covariance approximation. For cross-check studies, \texttt{diagonal} and \texttt{block} component models
are also available~\cite{Adye:2011gm}.

\subsection{Combination Model}
\label{sec:combination_model}

The total bin uncertainty combines all seven components in quadrature:
\begin{equation}
\Delta_{\mathrm{tot}}(j)=\sqrt{\Delta_{\mathrm{stat}}^2(j)
+ \Delta_{\mathrm{iter}}^2(j) + \Delta_{\mathrm{exp}}^2(j)
+ \Delta_{\mathrm{theory}}^2(j) + \Delta_{\mathrm{had}}^2(j)
+ \Delta_{\mathrm{ISR}}^2(j)}.
\end{equation}
The quadrature sum is used since the components originate from distinct sources and the covariance cross terms are expected to be small.

\subsection{Uncertainty Breakdown}
\label{sec:uncertainty_breakdown}

\begin{figure}[ht!]
\centering
\includegraphics[width=0.49\textwidth]{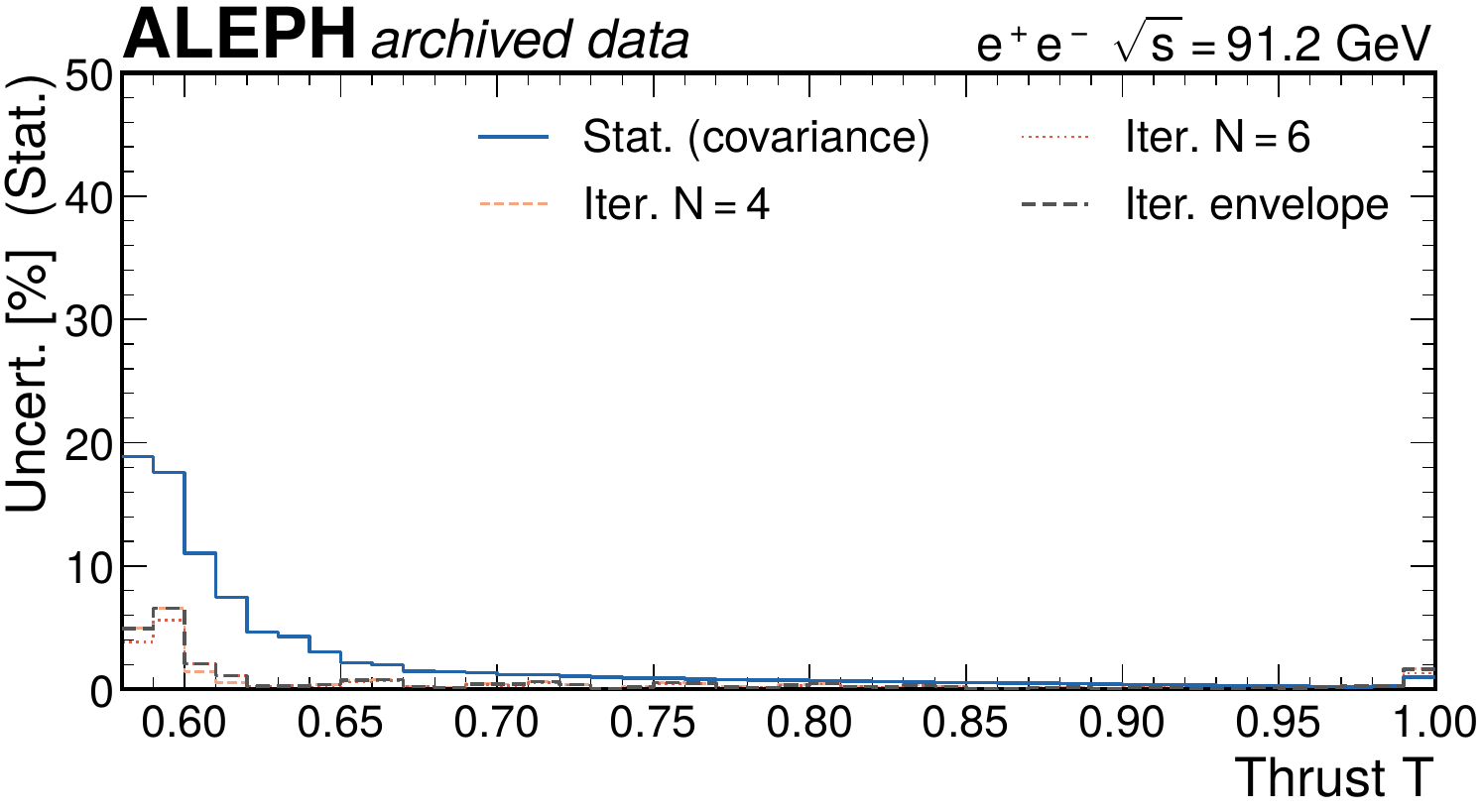}%
\hfill%
\includegraphics[width=0.49\textwidth]{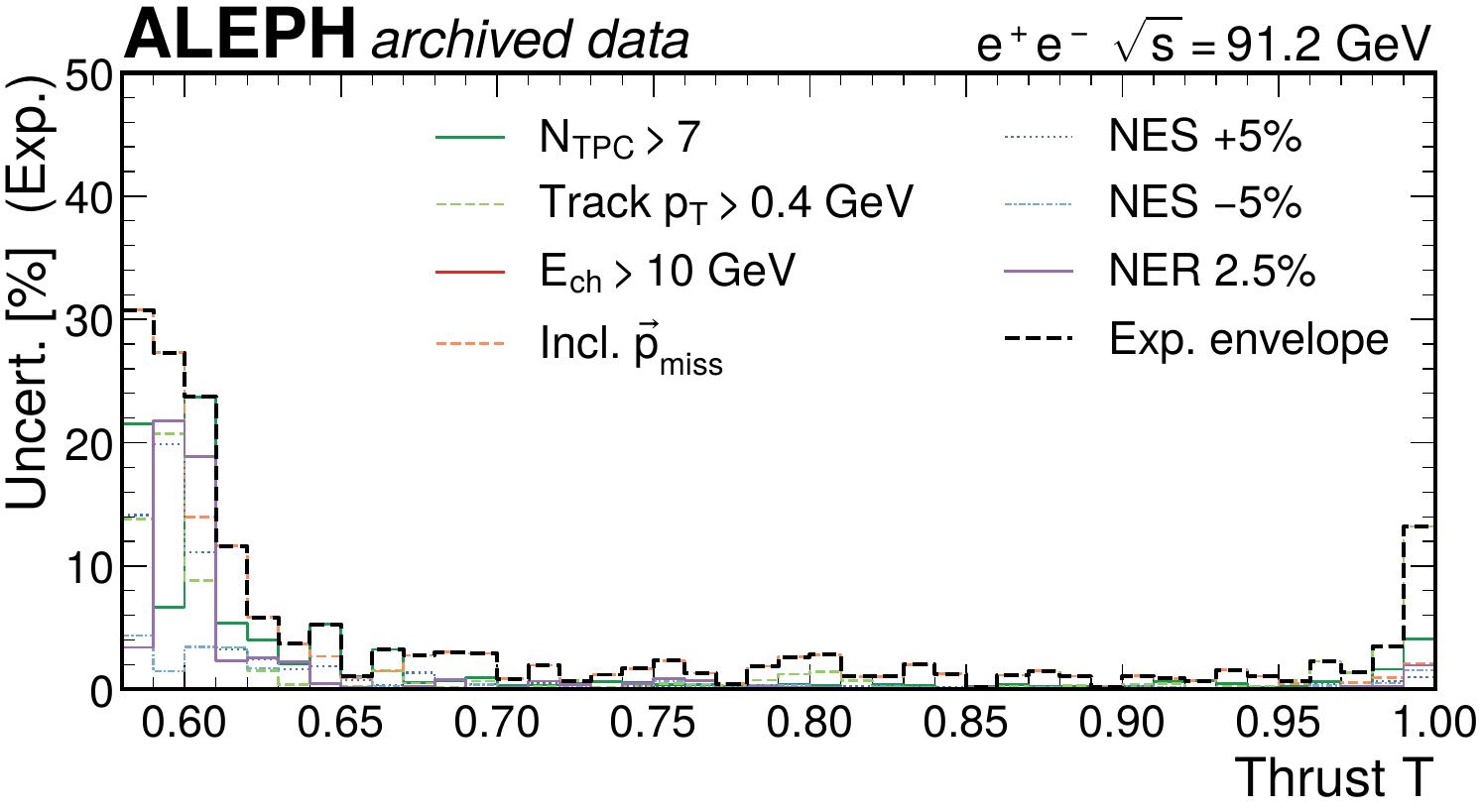}\\[2mm]
\includegraphics[width=0.49\textwidth]{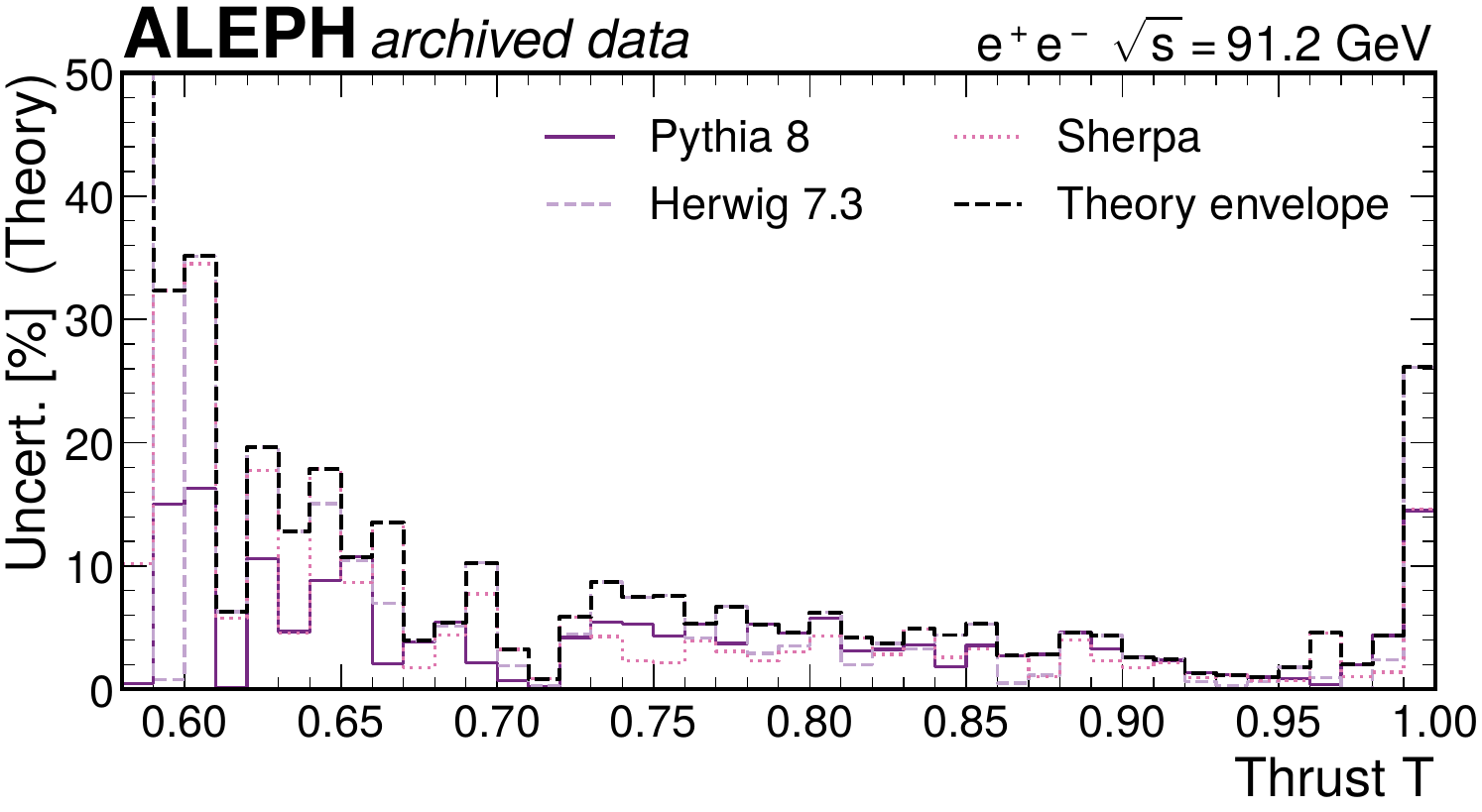}%
\hfill%
\includegraphics[width=0.49\textwidth]{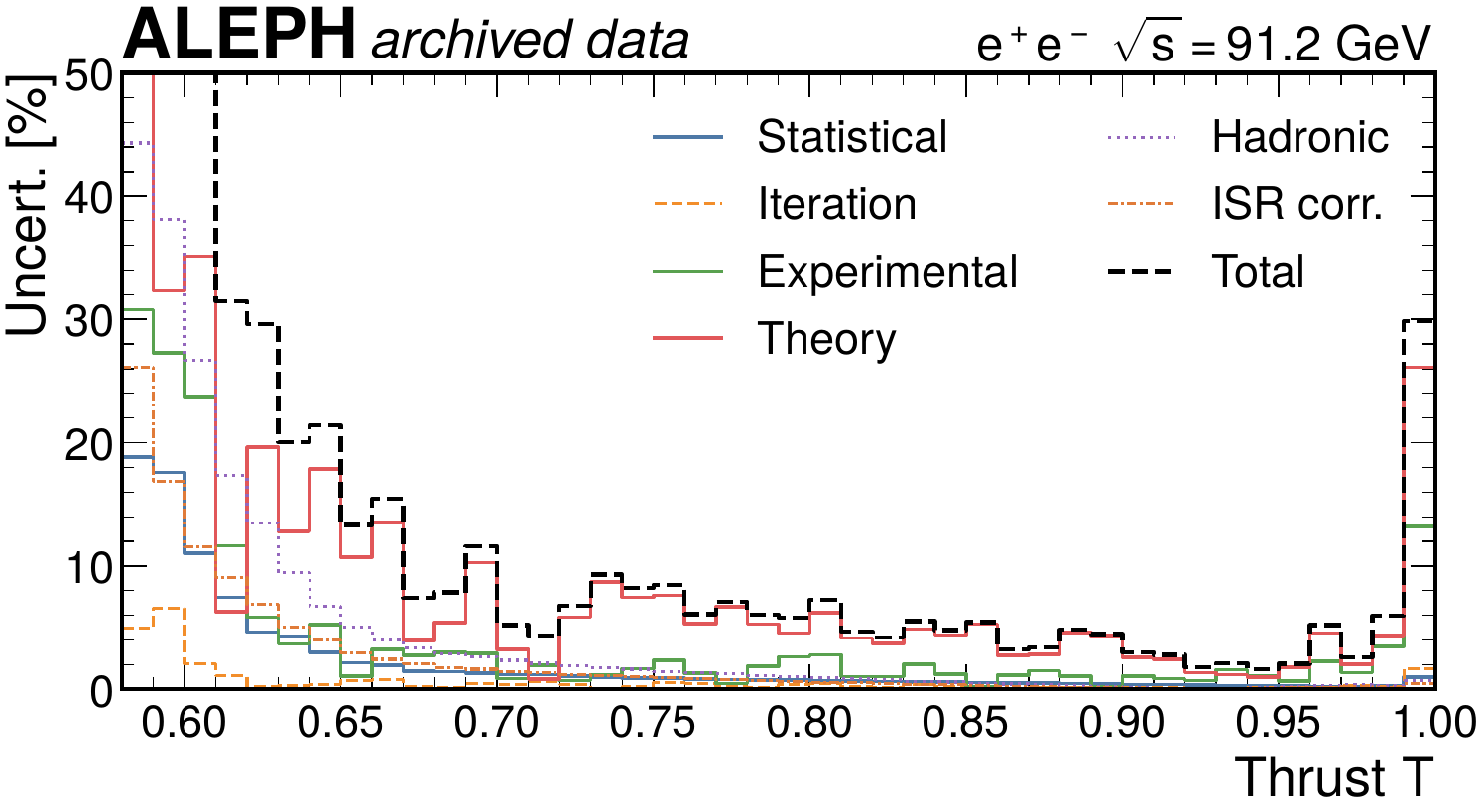}
\caption{Fractional uncertainties: (top left) statistical and unfolding-iteration components, (top right) experimental variation branches with group envelope, (bottom left) theory reweighting branches with theory envelope, (bottom right) all groups overlaid.
}
\label{fig:per_bin_uncertainty_grid}
\end{figure}

Figure~\ref{fig:per_bin_uncertainty_grid} shows the absolute fractional uncertainties. Theory is the dominant source across the central range, where the experimental envelope is smaller by roughly a factor of two. The statistical and iteration components remain sub-percent in the bulk.

\section{Results}
\label{sec:results}

\begin{figure}[ht!]
\centering
\includegraphics[width=0.85\textwidth]{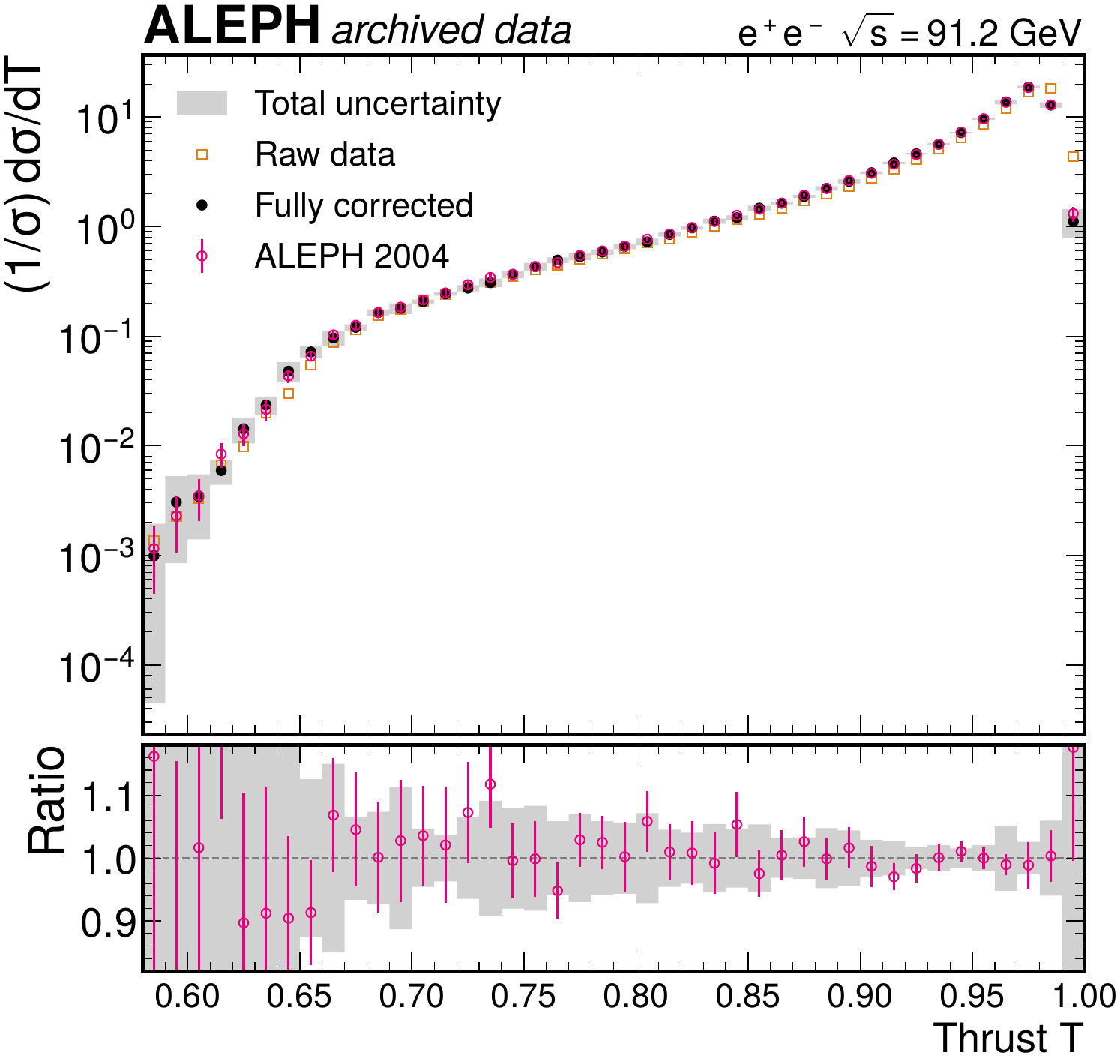}
\caption{Fully corrected thrust distribution with total uncertainty band compared to the published ALEPH 2004 result.}
\label{fig:final_thrust_result}
\end{figure}

Figure~\ref{fig:final_thrust_result} shows the unfolded spectrum with the total uncertainty band and comparison to the published ALEPH 2004 result~\cite{Heister:2003aj}. Agreement is good across the bulk of the spectrum ($0.62\le T<0.99$), with the largest discrepancy in the highest thrust bin. This is physically expected: the dijet limit is sensitive to MC modeling of soft hadrons, and the hadronic correction
$C_{\mathrm{had}}(j)$ and ISR correction $C_{\mathrm{ISR}}(j)$ carry growing uncertainties there. Potential dataset differences between the current analysis and the ALEPH 2004 result can also contribute at the endpoint.

\begin{figure}[ht!]
\centering
\sfpanel[0.32]{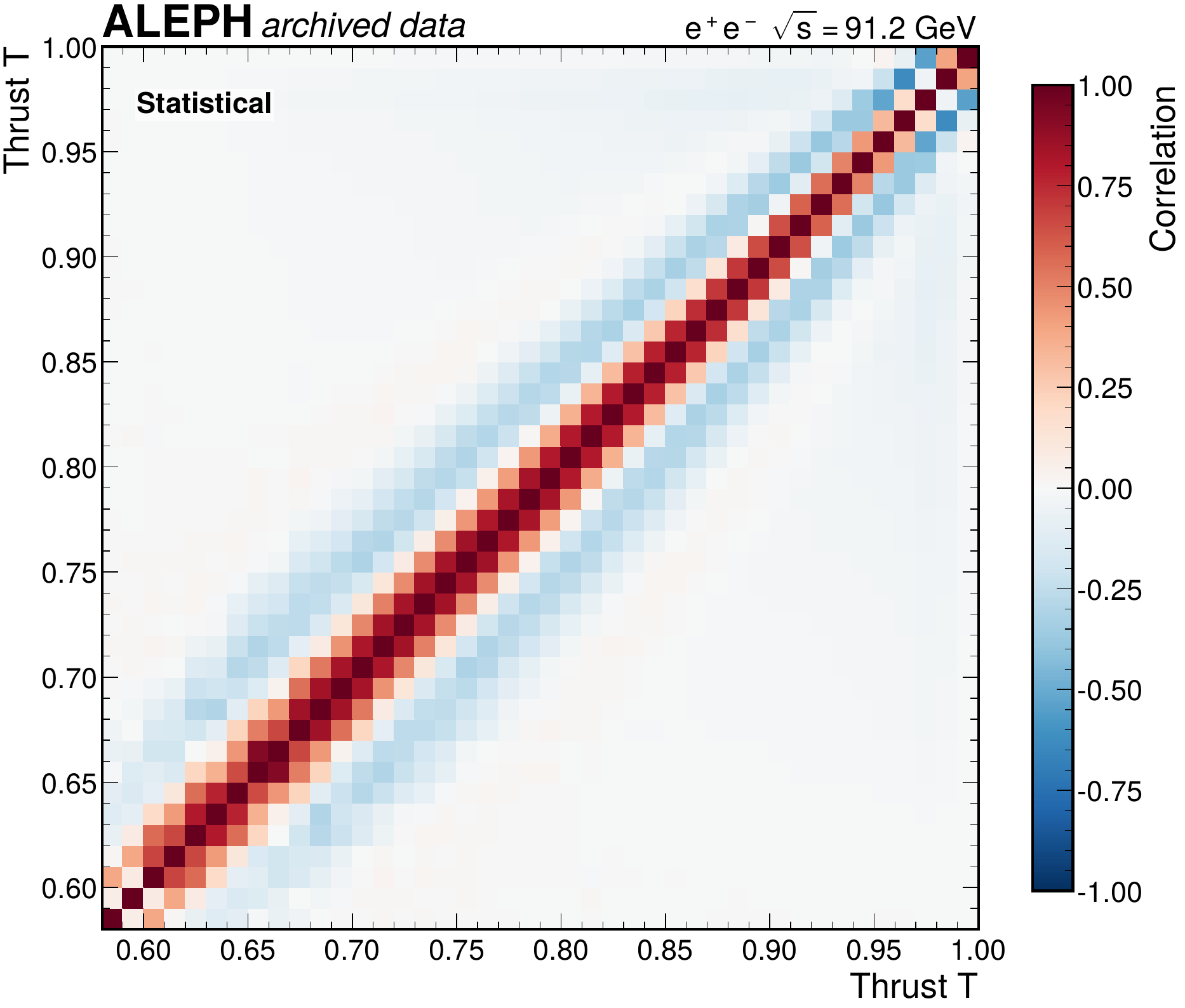}
\sfpanel[0.32]{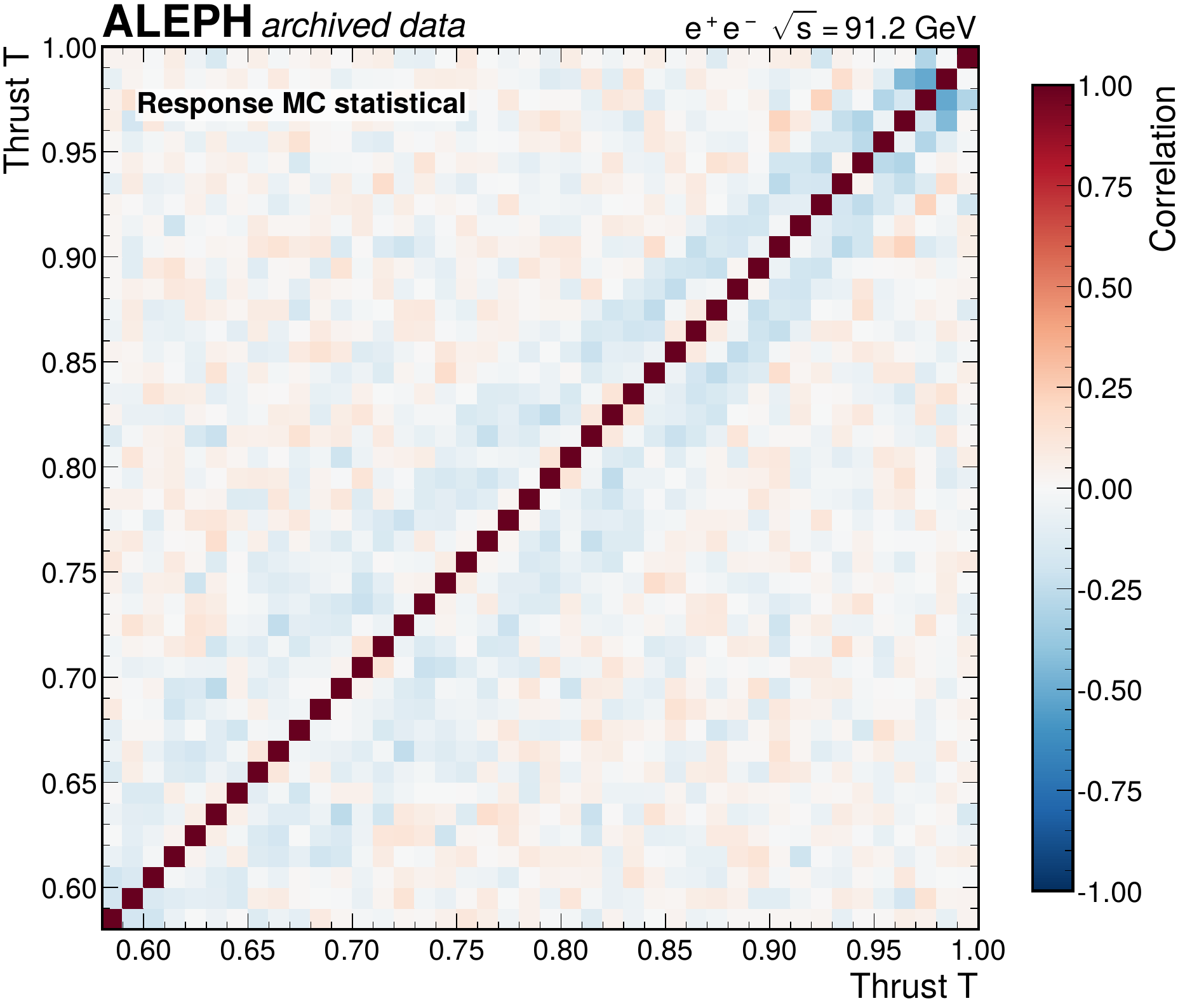}
\sfpanel[0.32]{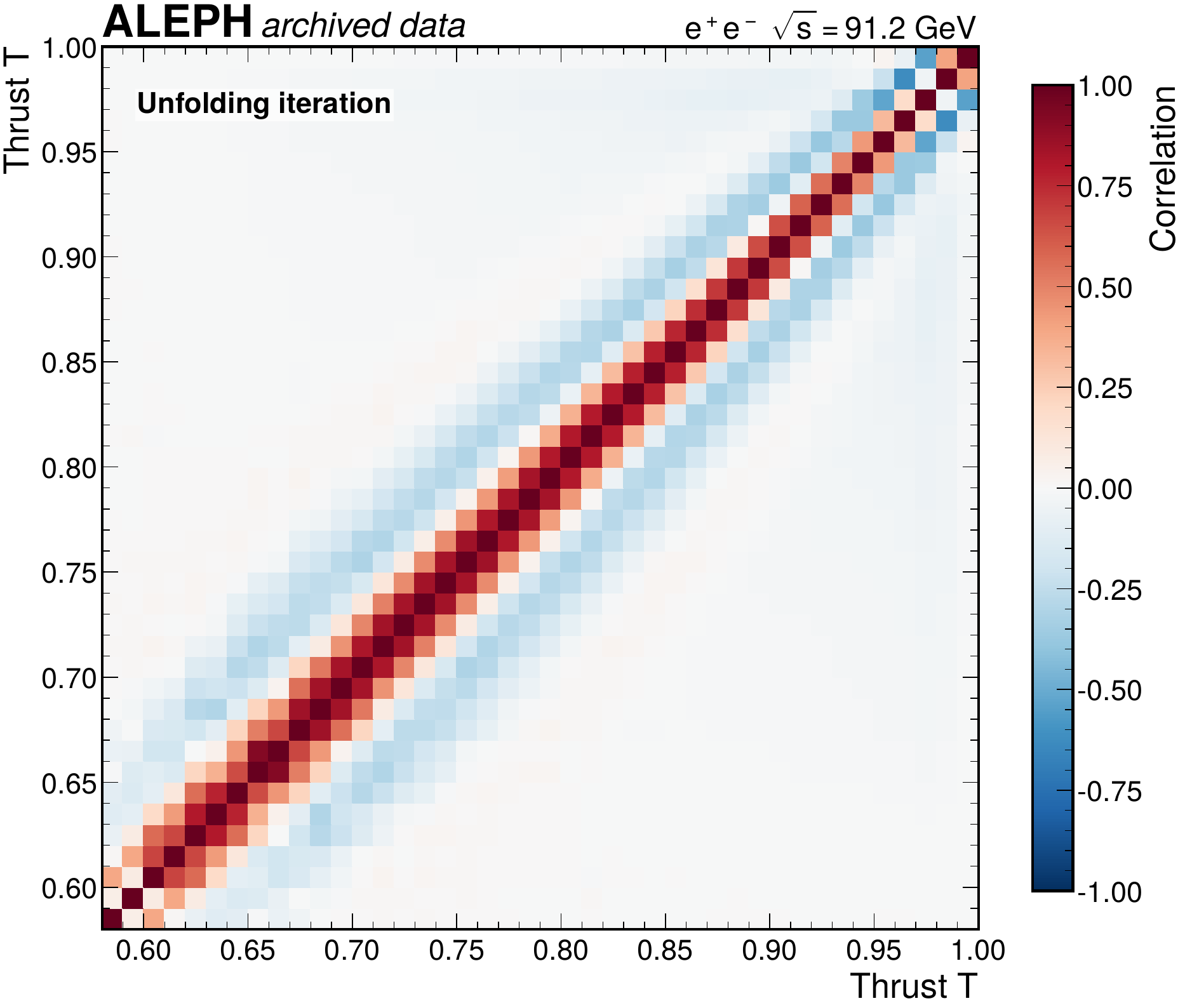} \hfill
\vspace{0.5em}
\sfpanel[0.32]{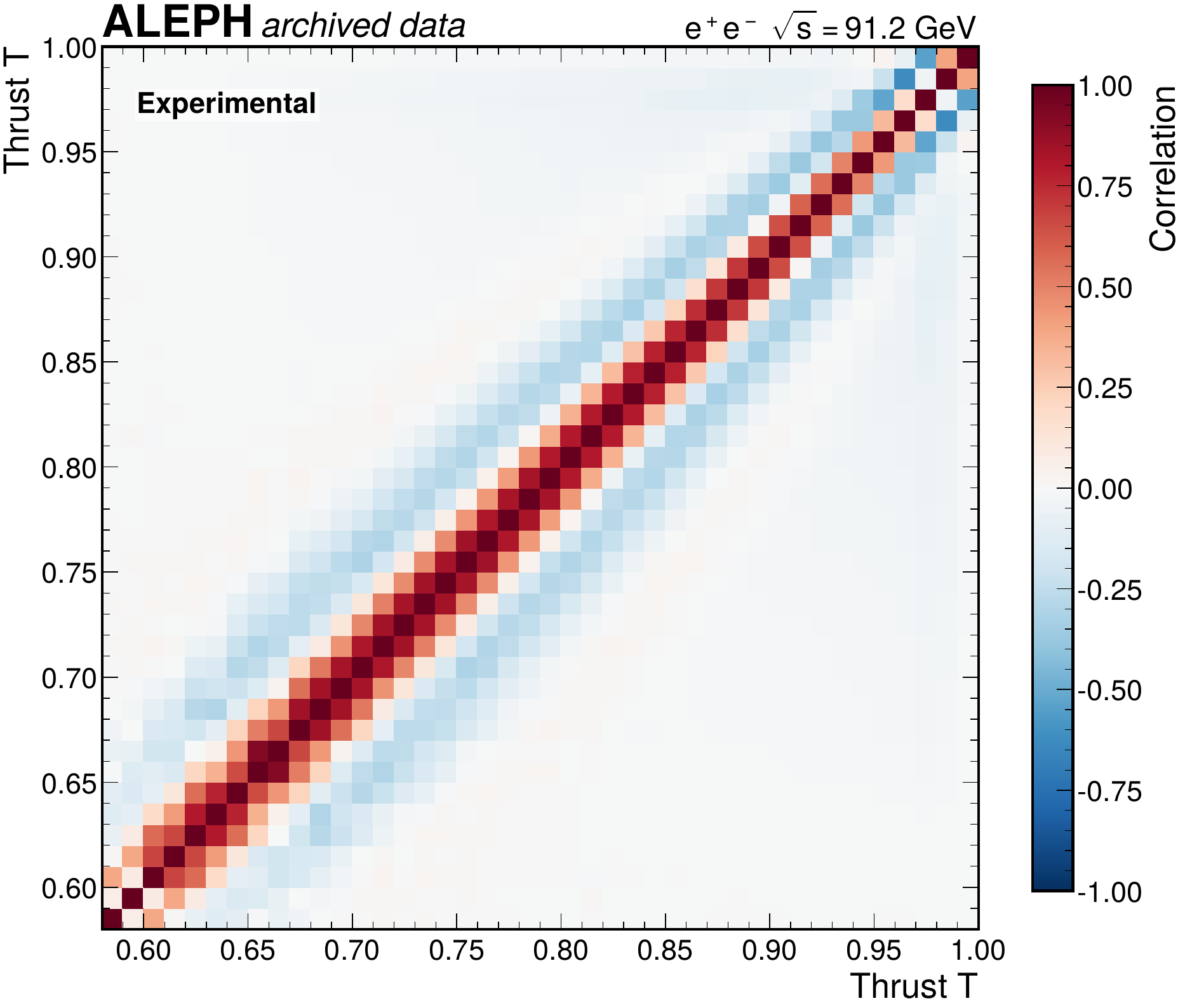}
\sfpanel[0.32]{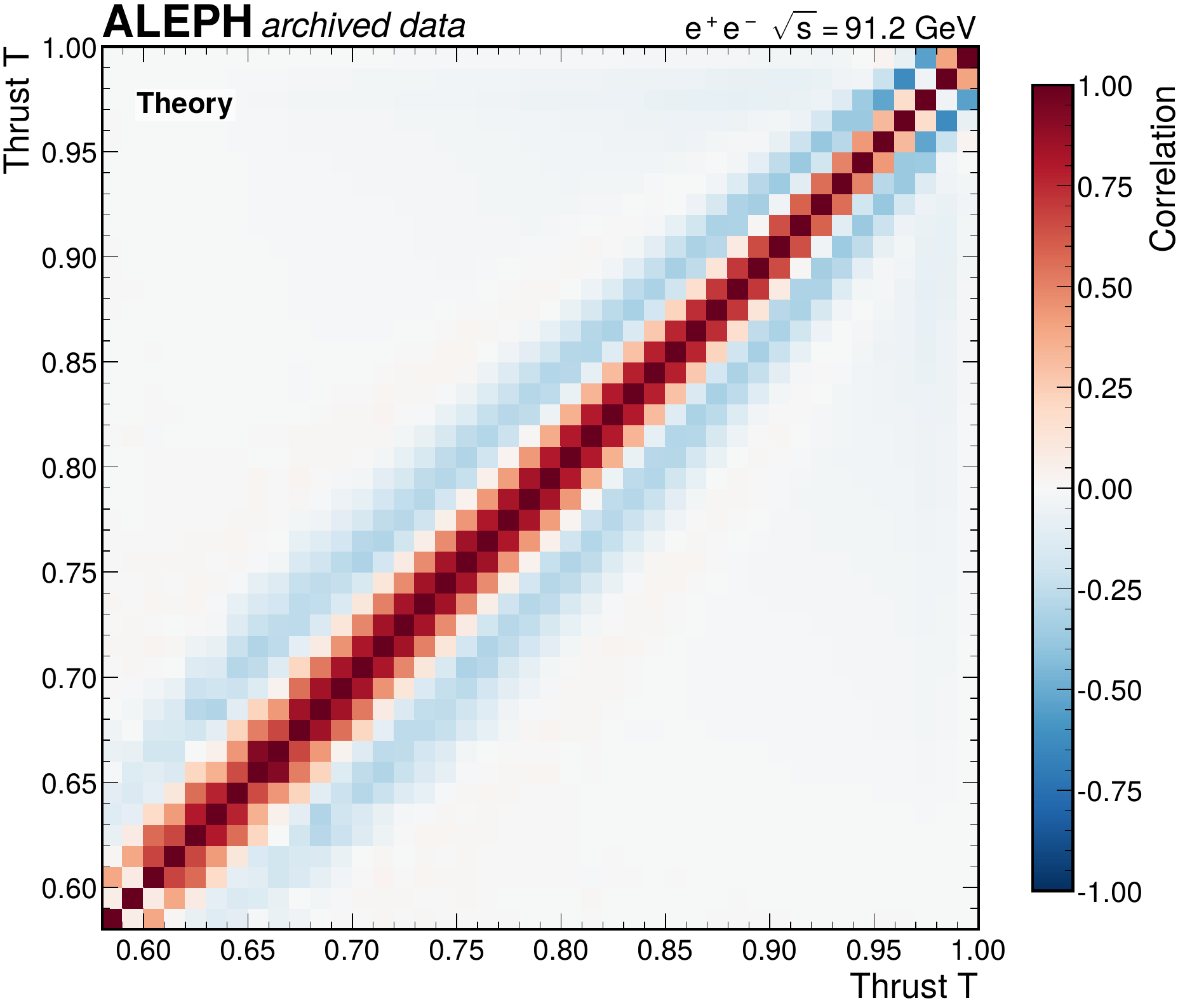}
\sfpanel[0.32]{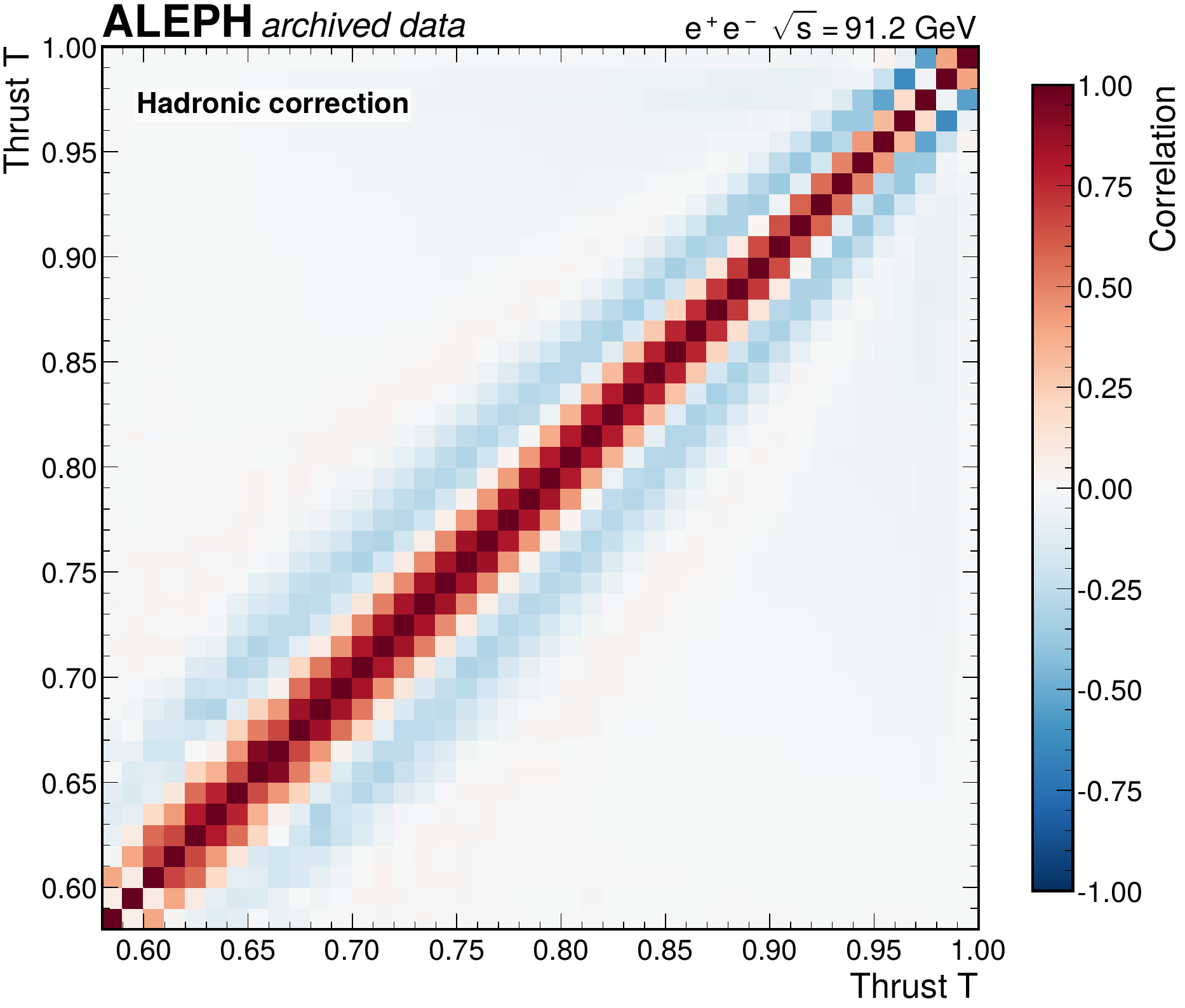} \hfill
\vspace{0.5em}
\sfpanel[0.32]{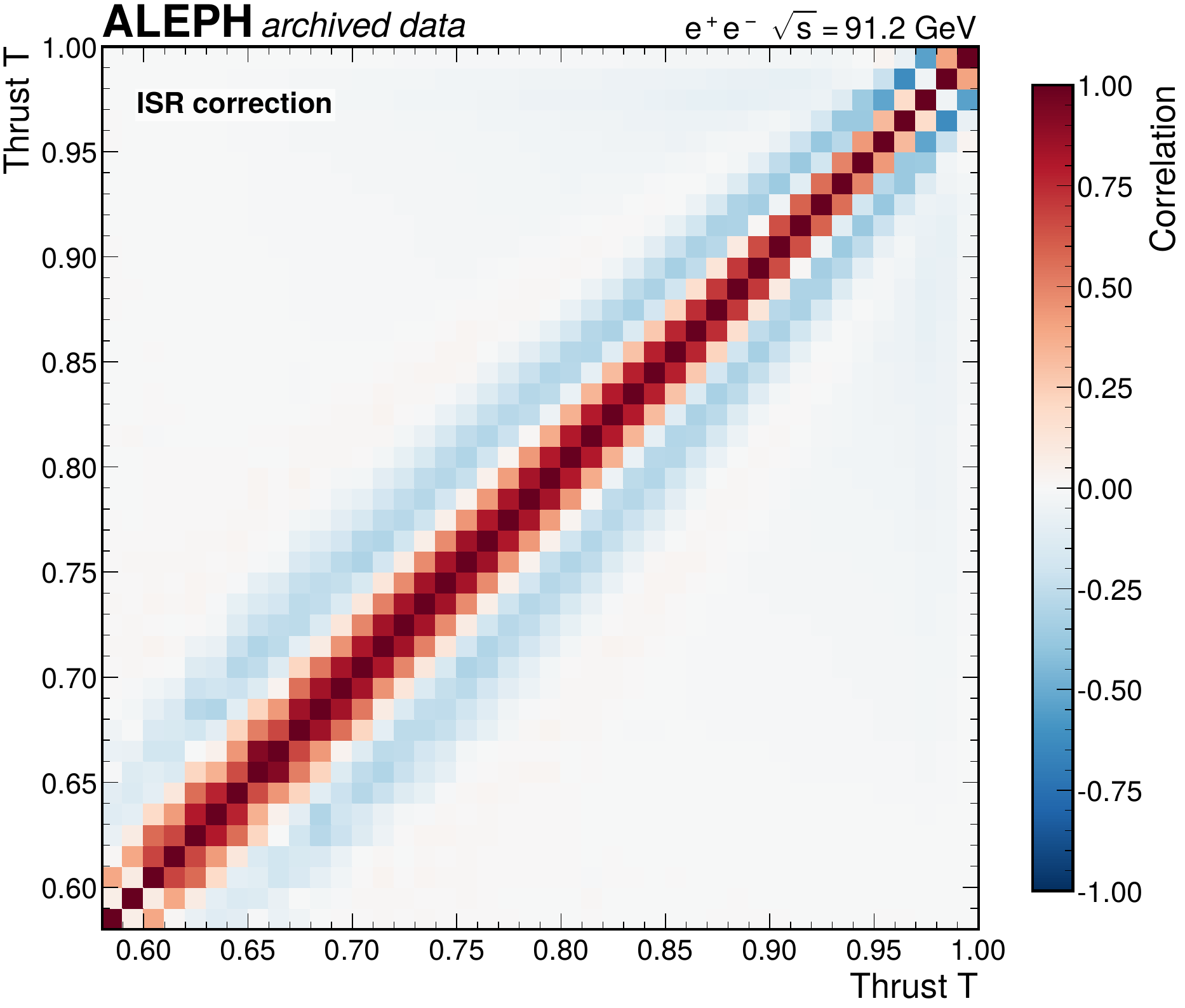}
\caption{Correlation matrices for the (a) data statistical, (b) response matrix MC statistical, (c) unfolding iteration, (d) experimental variation, (e) theory prior, (f) hadronic event selection correction, and (g) QED ISR correction components as indicated.}
\label{fig:corr_matrices_per_source}
\end{figure}

The per-component correlation matrices are derived from the individual component covariance matrices $\mathrm{Cov}^{(c)}_{ij}$ described in
Section~\ref{sec:stat_covariance}. For each component $c$, the correlation is 
\begin{equation}
\rho^{(c)}_{ij} = \frac{\mathrm{Cov}^{(c)}_{ij}}{\sqrt{\mathrm{Cov}^{(c)}_{ii}\,\mathrm{Cov}^{(c)}_{jj}}}.
\end{equation}
Figure~\ref{fig:corr_matrices_per_source} shows all seven components simultaneously, enabling direct comparison of their correlation structures. The components differ mainly in correlation length, and the ordering follows directly from whether the underlying variation is bin-local or coherent across the spectrum. The data statistical component exhibits the narrow correlation band characteristic of IBU: neighboring bins are positively correlated through the unfolding propagation, but the correlation falls off quickly with bin separation. The response matrix MC statistical component behaves the same way, since both originate in bin-local Poisson fluctuations spread over neighboring bins by the unfolding. The ISR correction is nearly diagonal, consistent with its sub-percent size and with the bin-local character of the MC statistical uncertainty on the $C_{\mathrm{ISR}}(j)$ ratio, and the hadronic correction is modestly correlated, commensurate with its ${\sim}2\%$ magnitude in the central range. The iteration component is somewhat broader, as the regularization responds coherently to low-frequency features of the spectrum rather than to individual bins. Broader still is the experimental component, driven by the neutral-energy response shifts (\texttt{nes\_up}, \texttt{nes\_down}), which displace the spectrum in the
same direction over a wide range of bins. The theory component has the strongest and most extended off-diagonal correlations: generator reweighting alters the parton-shower and hadronization description, and therefore the overall spectral shape, rather than any localized
region. It is the dominant source of off-diagonal structure.

\section{Conclusions}
\label{sec:conclusion}

AI agents have successfully carried out a complete thrust measurement using archived ALEPH 1994 data at $\sqrt{s}=91.2$~GeV under physicist guidance and collaboration. The analysis propagates uncertainty components arising from statistical, experimental, and theoretical sources and combines them in quadrature. The corresponding correlation matrices are derived from the covariance structure. The implementation of the ISR correction, following the prescription of both published ALEPH event-shape analyses~\cite{Barate:1996fi,Heister:2003aj}, brings the measured thrust distribution into good agreement with the published ALEPH 2004 result, with the only notable
discrepancy at the $T\to 1$ dijet endpoint. The AI-agent operating model demonstrated in this cycle, executable-workflow orchestration under physicist supervision with full
provenance tracking, provides a scalable framework for precision QCD as a testing ground for closed theory--experiment loops in fundamental physics.

\subsection*{Lessons from AI--Physicist Collaboration}

Carrying out this analysis has also produced practical insights into how AI agents and physicists can work together effectively. Our workflow followed a constrained human-in-the-loop cycle in which the physicist defined physics objectives, acceptance criteria, and interpretation authority, while the AI agent owned implementation and execution. Crucially, experienced physicists contribute domain context that cannot yet be reduced to formal specifications alone, including judgments about physically meaningful binning, acceptable regularization behavior, uncertainty communication, and when apparent anomalies warranted deeper investigation. A major future goal identified in this work is to translate this expert knowledge into explicit, code-based verification layers. 

From this experience, several recurring challenge classes emerged. Implicit physics priors, expert expectations about acceptable binning, regularization behavior, or uncertainty presentation, were often assumed rather than explicitly stated, requiring iterative clarification. Figure formatting and style choices (label offsets, legend placement, pad geometry) required many short feedback loops, with cosmetic shortcomings sometimes affecting interpretability and therefore physics conclusions. Synchronization across rapidly evolving scripts, outputs, and note text demanded continuous provenance refreshing. These lessons motivate creation of reusable tools, stronger configuration contracts, clearer channels for encoding expert heuristics into machine readable constraints in future iterations of this program, and new verification layers throughout the autonomous analysis flow.

\FloatBarrier
\section*{Author Contributions}
A.~Badea: conceptualization, primary analyzer, data curation, methodology, implementation, draft preparation. M.M., Y.-J.L.: data curation, review, feedback. C.M., A.~Baty, Y.C., H.B., G.M.I., J.Z., Y.-C.C., T.-A.S.: data curation. E.P.A.: Experimental collaboration under which all authors contribute significantly to data curation, physics analysis, draft preparation, and review.

\section*{Funding}
Y.C. and J.Z. are supported in part by the U.S. National Science Foundation under ID 2514008.
A.~Baty is supported by the U.S. Department of Energy, Grant No. DE-FG02-94ER40865.
H.B., Y.-C.C., G.M.I., C.M., Y.-J.L., and T.-A.S. are supported by the U.S. Department of Energy, Office of Science, Office of Nuclear Physics under grant Contract Number DE-SC0011088.
A.~Badea is supported by the Schmidt Sciences Foundation and Enrico Fermi Institute.

\section*{Data Availability Statement}
The data that support the findings of this study are available from the corresponding author upon reasonable request. The analysis code is made available through the Electron-Positron Alliance git organization: \url{https://github.com/orgs/Electron-Positron-Alliance}.

\section*{Acknowledgments}
The authors would like to thank Roberto Tenchini and Guenther Dissertori from the ALEPH Collaboration for their useful comments and suggestions on the use of ALEPH data.
We thank all the members of the Electron-Positron Alliance for their tremendous efforts to resurrect the LEP dataset, ultimately enabling the studies presented in this paper.

\section*{Conflicts of Interest}
The authors declare no conflict of interest.

\section*{Use of AI and AI-Assisted Technologies}
During the preparation of this work, the authors used Anthropic Claude and OpenAI GPT (ChatGPT and Codex) to carry out the experimental analysis and draft the paper. After using this tool/service, the authors reviewed and edited the content as needed and take full responsibility for the content of the published article.

\bibliographystyle{JHEP}
\bibliography{biblio}


\end{document}